\documentclass[a4paper,onecolumn,11pt,accepted=2024-12-04]{quantumarticle}
\pdfoutput=1
\usepackage[utf8]{inputenc}
\usepackage[english]{babel}
\usepackage[T1]{fontenc}
\usepackage{amsfonts}
\usepackage{amssymb}
\usepackage{mathtools}
\usepackage{amsmath}
\usepackage{amsthm}
\usepackage{hyperref}
\usepackage{cleveref}
\usepackage{tikz}
\usepackage{orcidlink}
\usepackage{lipsum}
\usepackage{braket}
\usepackage{bm}
\usepackage{appendix}
\usepackage{tikz}
\usepackage{algorithm}
\usepackage{algpseudocode}
\usepackage{natbib}
\usepackage{mathtools}
\newcommand{\verteq}{\rotatebox{90}{$\,=$}}
\newcommand{\equalto}[2]{\underset{\overset{\mkern4mu\verteq}{#2}}{#1}}

\crefname{thm}{Theorem}{Theorems}
\crefname{dfn}{Definition}{Definitions}
\crefname{rmk}{Remark}{Remarks}
\crefname{lem}{Lemma}{Lemmas}
\crefname{cor}{Corollary}{Corollaries}

\newtheorem{prop}{Proposition}
\newtheorem{dfn}{Definition}%[section]
\newtheorem{cor}{Corollary}

\theoremstyle{remark}
\newcommand{\ad}{\text{ad}}

\begin{document}

\title{Quantum simulation of time-dependent Hamiltonians via commutator-free quasi-Magnus operators}

\author{Pablo A. M. Casares \orcidlink{0000-0001-5500-9115}}
\email{pablo.casares@xanadu.ai}
\affiliation{Xanadu, Toronto, ON, M5G 2C8, Canada}%Lines break automatically or can be forced with \\

\author{Modjtaba Shokrian Zini}
\affiliation{Xanadu, Toronto, ON, M5G 2C8, Canada}

\author{Juan Miguel Arrazola \orcidlink{0000-0002-0619-9650}
}
\affiliation{Xanadu, Toronto, ON, M5G 2C8, Canada}
\maketitle

\begin{abstract}
  Hamiltonian simulation is arguably the most fundamental application of quantum computers. The Magnus operator is a popular method for time-dependent Hamiltonian simulation in computational mathematics, yet its usage requires the implementation of exponentials of commutators, which has previously made it unappealing for quantum computing.
The development of commutator-free quasi-Magnus operators (CFQMs) circumvents this obstacle, at the expense of a lack of provable global numeric error bounds.
In this work, we establish one such error bound for CFQM-based time-dependent quantum Hamiltonian simulation by carefully estimating the error of each step involved in their definition.
This allows us to compare its cost with the alternatives, and show that CFQMs are often the most efficient product-formula technique available by more than an order of magnitude.
As a result, we find that CFQMs may be particularly useful to simulate time-dependent Hamiltonians on early fault-tolerant quantum computers. 
\end{abstract}

\section{Introduction}

%Context and why it is interesting
%Simulating Hamiltonians is one of the most foundational applications of quantum computers. 
One of the most fundamental applications of quantum computing is simulating the time evolution of a quantum system. However, the time-dependent version has received comparatively less attention, even though it plays a key role in important problems including quantum control~\cite{nielsen2006optimal,pang2017optimal} and the interaction of electronic systems with external fields~\cite{alvermann2012numerical,auzinger2020photovoltaic,auzinger2022efficient}. It is also a core component of real-time electronic structure methods~\cite{goings2018real,li2020real}.

%Background results
%Suzuki -> product formula
%Dyson -> LCU
Quantum Hamiltonian simulation algorithms can be broadly grouped into two families of techniques: Linear Combination of Unitaries (LCU)~\cite{childs2012LCUs} and product formulas~\cite{childs2021theory}. There have been proposals in each of these families for the simulation of time-dependent Hamiltonians. Notably, the Dyson series in the LCU category provides an efficient, and in some regimes state-of-the-art, interaction-picture technique for this purpose~\cite{low2018hamiltonian,kieferova2019simulating,babbush2019quantum,su2021fault}. In the product formula category, prior arts include the Suzuki and continuous qDRIFT methods~\cite{wiebe2010higher,berry2020time}.

Here, we study yet another approach, the Magnus operator~\cite{magnus1954exponential}, which is an alternative to the explicit time-ordering of operators indicated by the Dyson series and Suzuki methods~\cite{magnus1954exponential,blanes2009magnus}, and enjoys popularity in the field of quantum control~\cite{ribeiro2017systematic,shapira2020theory,le2022analytic}. Recently, the Magnus operator has been adapted to its use under the LCU paradigm~\cite{an2022time,fang2024time}. However, it has received less attention in its product formula form as a quantum algorithm, perhaps due to the need to implement exponentials of commutators~\cite{childs2013product,chen2022efficient,bosse2024efficient,casas2024commutators}. Fortunately, the development over the past two decades of commutator-free quasi-Magnus operators (CFQMs) bypasses this necessity~\cite{blanes2006fourth,alvermann2011high,blanes2017high,blanes2009magnus}. CFQMs are commutator-free product formulas that approximate the Magnus operator evolution to a desired order.

A key advantage of most quantum algorithms for electronic Hamiltonian simulation is their ability to efficiently achieve a desired target error~\cite{childs2019nearly}. In contrast, classical algorithms implementing Magnus and quasi-Magnus operators suffer from an error that is less tightly controlled because they often need to approximate the computation of each exponential, relying for instance on Krylov subspace or Lanczos methods~\cite{bader2018exponential,iserles2018magnus}; though there are exceptions~\cite{celledoni2000approximating}. 

Conversely, classical algorithms have the advantage of being able to efficiently read, prepare, or copy any state in memory. As such, the known methods that bound the algorithmic error are to the best of our knowledge `a posteriori' or adaptative, that is, they leverage knowledge of an approximate solution to compute the error incurred at each step~\cite{wensch2004solution,kormann2008accurate,kormann2011global,auzinger2019posteriori,auzinger2019symmetrized,auzinger2021adaptive}. Unfortunately, this strategy is much less amenable to quantum computing, where preparing and reading out states is costly and requires recomputing the state preparation and time evolution up to the point of interest~\cite{ikeda2023trotter24,zhao2023adaptive}. Additionally, the convergence of CFQMs has been studied previously, including for unbounded operators, and partial results have been reported~\cite{blanes2018convergence,blanes2020convergence,ikeda2023minimum,chen2023quantum}.
However, as far as we are aware, no explicit `a priori' error bounds were derived for CFQMs solving the time-dependent Schrödinger equation. This left unanswered the question of the largest time step that would guarantee a desired error when using them in a quantum algorithm.

Here, we fill this gap by enabling the use of CFQMs in quantum computing with the usual guarantees. Specifically, we bound the error incurred by using CFQMs to simulate bounded time-dependent Hamiltonians, whose Taylor series coefficients can also be bounded.
We compute the error incurred at each approximation used in the derivation of CFQMs, employing carefully chosen expressions and exploiting symmetries along the way.
The result is a set of quantum algorithms with similar asymptotic complexity as the Suzuki method~\cite{wiebe2010higher}, but better constant factor overhead for at least some classes of Hamiltonians, including time-dependent Heisenberg models, see~\cref{fig:Magnus_derivation}.

The paper is structured as follows. First, we present a succinct derivation of CFQMs, which is instrumental for our error bounds. Then, in \cref{sec:error} we analyze each of the approximations involved in their development and provide error bounds for each. In \cref{sec:results}, we compare CFQMs with previous works and provide numerical estimations for the simulation error of time-dependent Heisenberg Hamiltonians. We conclude with a summary of the results and a discussion on future works, as well as the potential of this technique in the early fault-tolerant regime. Details of our analysis can be found in the appendices. \cref{app:Comparison} gives a more detailed comparison of CFQMs performance against other time-dependent Hamiltonian simulation methods. \cref{app:proofs} contains the rigorous proofs, \cref{app:split_operator} evaluates specific extensions avoiding usage of Trotter product formulas for Hamiltonians composed of two self-commuting and fast-forwardable terms, and lastly, \cref{app:Multi-product formulas} discusses how CFQMs may be used in combination with multi-product formulas~\cite{low2019well}.

\section{Magnus and commutator-free quasi-Magnus operators}

The development of the Magnus operator starts with the following observation: in the time-independent Schrödinger equation, the operator $e^{-iHt}$ plays the role of the time propagator for a time-independent Hamiltonian $H$. Therefore, one may wonder if there is an operator $\Omega(t)$ such that $e^{\Omega(t)}$ is a propagator of the time-dependent Schrödinger equation
\begin{equation}
    \partial_t \psi(t) = -iH(t)\psi(t) =: A(t) \psi(t).
\end{equation}
The solution is the \textit{Magnus operator}~\cite{magnus1954exponential}, 
\begin{equation}\label{eq:Magnus_operator}
\Omega(t_0, t_0+\Delta t) = \sum_{n=1}^\infty\Omega_{(n)}(t_0, t_0+\Delta t),
\end{equation}
with 
\begin{equation}\label{eq:Omega_1}
\Omega_{(1)}(t_0, t_0+\Delta t) = \int_{t_0}^{t_0+\Delta t} d\tau A(\tau)
\end{equation}
and higher orders defined inductively as~\cite{blanes2009magnus}
\begin{align}\label{eq:Omega_n}
    &\Omega_{(n)}(t_0, t_0+\Delta t) = \sum_{j=1}^{n-1}\frac{b_j}{j!}\sum_{\substack{k_1 + \ldots + k_{j} = n-1\\ k_l\geq 1,\forall l\in\{1,\ldots,j\}}}\left(\int_{t_0}^{t_0+\Delta t} \ad_{\Omega_{(k_1)}(\tau)}\ldots \ad_{\Omega_{(k_j)}(\tau)} A(\tau) d\tau\right),
\end{align}
where $\ad_B(C) := [B,C]$ and the $b_j$'s are the Bernoulli numbers appearing as coefficients in the Taylor series of $\frac{z}{e^z-1}$. %We may use the notation $\Omega(t_0, t_0+t)$ to indicate time evolution starting at $t_0$ instead of $0$.
The series in~\cref{eq:Magnus_operator} does not always converge. There is instead a finite radius of convergence over $t$, and thus, to obtain a global solution, we have to split the total time evolution into segments. The operator $\Omega(t_0,t_0+\Delta t)$ has been proven to be absolutely convergent, and thus well-defined, if
%\cite[Eq. 22]{blanes1998magnus}
$
    \int_{t_0}^{t_0+\Delta t} \|A(\tau)\|d\tau \leq 1.086869,
$
or if $\max_{\tau\in [t_0,t_0+ \Delta t]}\|A(\tau)\| \Delta t < 2$ \cite{blanes1998magnus,moan2001convergence}.
\begin{figure*}
    \centering
    \scalebox{0.81}{
    \begin{tikzpicture}[node distance=2cm, auto]

   % Node with more than one line and a name
  %\node[align=center, draw, rounded corners] (A) at (0,0) {$A(t)$};

  \node[align=center, draw, rounded corners] (A_Tay) at (3,0) {Hamiltonian \\$A(t) = \sum_n a_n (t-t_{\frac{1}{2}})^{n}$};

 %\draw[->] (A) -- (A_Tay);

  \node[align=center, draw, rounded corners] (Magnus) at (0,1.5) {Magnus \\ operator \\ $\Omega(h)$};

  \node[align=center, draw, rounded corners] (Magnus_Tay) at (3,1.5) {Magnus operator \\ as a function of $\alpha_n$};

  \draw[->] (A_Tay) -- (Magnus_Tay);
  \draw[->] (Magnus) -- (Magnus_Tay);

  \node[align=center, draw, rounded corners] (Omega2s) at (7,1.5) {$\Omega^{[2s]}(h)$ as a\\ function of $\alpha_j$,\\
 $j = 1,\ldots,s$};

  \draw[->, dotted] (Magnus_Tay) -- (Omega2s);

  \node[align=center, draw, rounded corners] (TildeOmega2s) at (11,1.5) {$\tilde{\Omega}^{[2s]}(h)$ as a  \\ function of $A^{(g)}(h)$,\\
$g = 0,\ldots,s-1$};

  \draw[->, dotted] (Omega2s) -- (TildeOmega2s);

  \node[align=center, draw, rounded corners] (BarOmega2s) at (15.3,1.5) {$\overline{\Omega}^{[2s]}(h)$ as a \\ function of $A(\tau_k) = A_k$\\
  $k = 1,\ldots,s$};

  \draw[->, dashdotted] (TildeOmega2s) -- (BarOmega2s);

    \end{tikzpicture}
    }
\caption{\label{fig:Magnus_derivation_simple}Flowchart indicating how the Magnus operator can be expressed using several basis sets. The dash-dotted line indicates equality up to $O(h^{2s})$; whereas the dotted ones indicate a formal identification between two Lie algebras indexed by $O(h^p)$ (expanded by $\{\alpha_j\}$) and without indexing (expanded by $\{A_k\}$ or $\{A^{(g)}(h)\}$). To transform $\Omega^{[2s]}$ into $\tilde{\Omega}^{[2s]}$, the matrix $(T^{[2s]})^{-1}$ is used, see e.g.~\cref{eq:T2^-1_example,eq:T3^-1_example}. To transform $\tilde{\Omega}^{[2s]}$ into $\overline{\Omega}^{[2s]}$ the quadrature matrix is used, see~\cref{eq:Gauss_Legendre_quadrature}.}
\end{figure*}
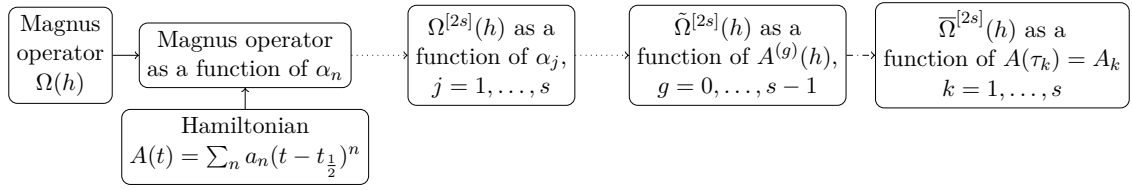
Implementing the Magnus operator in its original form is complicated, as it requires using exponentials of integrals of commutators. While there exist quantum algorithms to implement such exponentials \cite{childs2013product,chen2022efficient}, we instead resort to commutator-free operators that approximate $\Omega(t_0,t_0+\Delta t)$.

\subsection{Lie algebras and basis changes}

We denote by $h=\Delta t$ the length of the segments needed to break down the evolution. Thus, the Magnus operators we consider are of the form $\Omega(t_0,t_0+h)$. From here onward, we may drop $t_0$ and abuse $\Omega(h)$ as a substitute for $\Omega(t_0,t_0+h)$, as our analysis and equations identically apply to any $t_0$. The same applies to future functions which are similarly defined on the segments.

To derive a CFQM, we substitute the Hamiltonian in \cref{eq:Omega_n} by its Taylor series around the midpoint $t_{\frac{1}{2}} = t_0 + \frac{h}{2}$,
\begin{equation}\label{eq:Taylor}
    A(t) = \sum_{n= 0}^\infty a_n (t-t_{\frac{1}{2}})^n, \qquad a_n = \left.\frac{1}{n!}\frac{d^n A(t)}{dt^n}\right|_{t=t_{\frac{1}{2}}}.
\end{equation}
Using the Taylor expansion allows us to separate the time dependency of the operator from the matrix, and analytically compute the integrals in~\cref{eq:Omega_1,eq:Omega_n}.

A Lie algebra $\mathfrak{g}$ is defined as a vector space with a binary operation $[\cdot, \cdot]$ satisfying bilinearity, the alternating property, and the Jacobi identity. Its basis is the basis of $\mathfrak{g}$ understood as a vector space~\cite[Appendix A.3]{blanes2017concise}. %~\cite[Appendix A.3]{blanes2017concise}%~\cite[Definition 3.27]{hall2000elementary}. A basis of the Lie algebra $\mathfrak{g}$
Let $\alpha_n := a_{n-1}h^n$. The $\alpha_n$'s form the basis of a Lie algebra indexed by their $O(h^n)$ order~\cite{munthe1999computations,blanes2009magnus}.
This property makes them a good basis set to simplify the structure of $\Omega(t_0, t_0+h)$, reducing the number of commutators~\cite[Section 4b]{munthe1999computations}. %The $\{\alpha_j\}$ represent the first of three basis sets that we will use to manipulate the Magnus operator, see~\cref{fig:Magnus_derivation_simple}. 
%We make two observations. First, 
%imposing the time anti-symmetry $\exp(\Omega(-t)) = \exp(\Omega(t))^{-1} = \exp(-\Omega(t))$,
On the other hand, the function $\Omega\left( t_{\frac{1}{2}} - \frac{h}{2}, t_{\frac{1}{2}} + \frac{h}{2}\right) = -\Omega\left( t_{\frac{1}{2}} + \frac{h}{2}, t_{\frac{1}{2}} - \frac{h}{2}\right)$ is antisymmetric around $t_{\frac{1}{2}}$. As a result, only odd powers of $h$ will appear in the expansion of $\Omega(h)$ around $t_{\frac{1}{2}}$~\cite{blanes2000improved}. 
 
Next, we define the formal representations $\Omega^{[2s]}(h)$ of the $2s$-th order expansion of $\Omega(h)$ in the Lie algebra generated by $\alpha_j$ with $j = 1,\ldots,s$. In other words, we (i) substitute~\cref{eq:Taylor} in~\cref{eq:Magnus_operator,eq:Omega_1,eq:Omega_n}, (ii) perform the integrations, (iii) write the expression in terms of the $\alpha_j$'s, and (iv) take only those for which $j\leq s$~\cite{blanes2006fourth}.
%We do so by taking the terms involving those $\alpha_j$'s in the Taylor expansion of the Magnus operator. 
For instance,
\begin{align}\label{eq:Omega2_1generators}
    \Omega^{[2]}(h) &= \alpha_1,\\
\label{eq:Omega4_2generators}
    \Omega^{[4]}(h) &= \alpha_1 - \frac{1}{12}[1,2],\\
    \label{eq:Omega6_3generators}
    \Omega^{[6]}(h) &= \alpha_1 - \frac{1}{12}[1,2] +\frac{1}{12}\alpha_3 + \frac{1}{240}[2,3]+ \frac{1}{360}[1,1,3] -\frac{1}{240}[2,1,2]+\frac{1}{720}[1,1,1,2],
\end{align}
where $[j_1,\ldots,j_k]$ is a compact notation for $[\alpha_{j_1},[\ldots[\alpha_{j_{k-1}}, \alpha_{j_k}]]]$, of order $O(h^{j_1+\ldots+j_k})$~\cite{blanes2009magnus}. We may omit the argument $h$ from $\Omega^{[2s]}$ and other operators when it is clear from context. These transformations constitute steps~\ref{alg:step_Magnus_Taylor} and~\ref{alg:step_Magnus_Taylor_trunc} in the construction of CFQMs, described in ~\cref{alg:Steps_CFQMs}.

Note that the term $\frac{1}{12}\alpha_3$ does not appear in $\Omega^{[4]}(h)$, even though it is an $O(h^3)$ term. Similarly, $\frac{1}{80}\alpha_5$ and $-\frac{1}{80}[1,4]$ will be missing from $\Omega^{[6]}(h)$ even though they are $O(h^5)$. Therefore, these should be viewed as formal expressions, not an approximation up to $O(h^{2s+1})$. 

There is a way to recover the missing terms and make this approximation by using a related Lie algebra; it involves numerically approximating the integrals at the \textit{quadrature nodes} $c_k$ of each time interval, such as those prescribed by the Gauss-Legendre method. Evaluating the Hamiltonian at $s$ many quadrature nodes, e.g. $A_k := A(c_k\frac{h}{2}+t_{\frac{1}{2}})$ with $c_k\in[-1,1]$, allows us to obtain a $2s$-order approximation, see~\cite[Corollary 3.3]{iserles1999solution} and~\cite{alvermann2011high,blanes2017high,blanes2018time}.
%It is possible to identify the two Lie algebras generated by $\{A_k\}$ and $\{\alpha_j\}$ such that the number of commutators necessary to approximate $\Omega(h)$ is greatly reduced~\cite[Section 4b]{munthe1999computations}.

However, before obtaining the approximation in terms of $\{A_k\}$, we need to introduce an intermediate set of variables, the \textit{univariate integrals},
\begin{equation}
\begin{split}
\label{eq:A^i_integrals}
    A^{(g)}(h) &:= \frac{1}{h^g}\int_0^h (t-t_{\frac{1}{2}})^g A(t) dt = \frac{1}{h^g}\int_{-h/2}^{h/2} t^g A(t+t_{\frac{1}{2}}) dt\\
    &= \sum_{j = 0}^\infty  \frac{1-(-1)^{g+j+1}}{(g+j+1)2^{g+j+1}} \alpha_{j+1} =: \sum_{j=1}^\infty T_{g,j}\alpha_{j}.
\end{split}
\end{equation}
Note the change of variable $j+1\rightarrow j$ in the last equality. 
Consequently, $A^{(0)}$ now reads
\begin{align}\label{eq:A^0_integral}
    A^{(0)}(h) &= \alpha_1 + \frac{1}{12} \alpha_3 + \frac{1}{80}\alpha_5 +\ldots
\end{align}
Let $T^{(s)}$ be the matrix with elements $T_{g,j}$ defined in~\cref{eq:A^i_integrals} with $g = 0,\ldots,s-1$ and $j = 1,\ldots,s$.
Since $T^{(s)}$ arises from a (positive semi-definite) inner product between linearly independent monomials $t^{j}$, it is invertible~\cite{Yuan20204InvertibilityT}. Consequently, univariate integrals similarly form the basis of a Lie algebra~\cite{blanes2006fourth}. We now define $\tilde{\Omega}^{[2s]}(h)$ by formally rewriting $\Omega^{[2s]}(h)$ in terms of the univariate integrals $\{A^{(g)}\}$ with $g=0,\ldots,s-1$ using $(T^{(s)})^{-1}$. For example,
\begin{align}\label{eq:T2^-1_example}
    \begin{pmatrix}
        \alpha_1\\
        \alpha_2
    \end{pmatrix}
    = 
    \begin{pmatrix}
        1 & 0\\
        0 & 12
    \end{pmatrix}
    \begin{pmatrix}
        A^{(0)}\\
        A^{(1)}
    \end{pmatrix}, &\qquad s = 2; \\
    \label{eq:T3^-1_example}
     \begin{pmatrix}
        \alpha_1\\
        \alpha_2 \\
        \alpha_3
    \end{pmatrix}
    = 
    \begin{pmatrix}
        \frac{9}{4} & 0 & 15\\
        0 & 12 & 0\\
        -15 & 0 & 180
    \end{pmatrix}
    \begin{pmatrix}
        A^{(0)}\\
        A^{(1)}\\
        A^{(2)}
    \end{pmatrix}, &\qquad s = 3.
\end{align}
Thus, substituting in~\cref{eq:Omega4_2generators,eq:Omega6_3generators}, we obtain~\cite{blanes2006fourth}.
\begin{align}
    \tilde{\Omega}^{[4]}(h) &= A^{(0)}(h) + [A^{(1)}(h), A^{(0)}(h)],\label{eq:tilde_Omega_4}\\
    \label{eq:tilde_Omega_6}
    \tilde{\Omega}^{[6]}(h) &= A^{(0)}(h)+\left[A^{(1)}(h), \frac{3}{2}A^{(0)}(h)-6A^{(2)}(h)\right] + \frac{1}{2}[A^{(0)}(h), A^{(0)}(h), A^{(2)}(h)]\\
   & +\frac{3}{5}[A^{(1)}(h), A^{(1)}(h), A^{(0)}(h)] + \frac{1}{60}[A^{(0)}(h), A^{(0)}(h), A^{(0)}(h), A^{(1)}(h)].\nonumber
\end{align}
We can now recover all the missing terms in \cref{eq:Omega2_1generators,eq:Omega4_2generators,eq:Omega6_3generators}, by fully expanding $\tilde{\Omega}^{[2s]}(h)$ in terms of all $\alpha_j$'s using \cref{eq:A^i_integrals}~\cite{blanes2000improved,blanes2009magnus}. For instance, using~\cref{eq:A^0_integral} we see $\tilde{\Omega}^{[4]}(h)$ includes $\frac{1}{12}\alpha_3$ (as well as $\frac{1}{80}\alpha_5$) missing in \cref{eq:Omega4_2generators}. 

Finally, one can use a quadrature method, for instance Gauss-Legendre, to approximate such integrals. Given that we have recovered all the missing $O(h^{2s})$ terms, we conclude with an $O(h^{2s+1})$-accurate expression of the Magnus operator, $\overline{\Omega}^{[2s]}(h)$, as a function of the Hamiltonian evaluated at the $s$ quadrature nodes of each time segment, $\{A_k\}$. The transformations described in this section are schematically depicted in~\cref{fig:Magnus_derivation_simple} and the upper row of~\cref{fig:Magnus_derivation} in~\cref{app:Comparison}. 

\subsection{Commutator-free quasi-Magnus operators}

We harness the machinery introduced above to define commutator-free operators that approximate the Magnus operator to the desired order. One may use the ansatz~\cite{blanes2006fourth}
\begin{equation}
\label{eq:CF_definition_integrals}
\begin{split}
    \tilde{U}_{m}^{[2s]}(h) &:= \prod_{i=1}^m \exp\left(\sum_{g=0}^{s-1} y_{i,g} A^{(g)}(h)\right) = \exp(\tilde{\Omega}^{[2s]}(h)) + O(h^{2s+1}),
\end{split}
\end{equation}
where $y_{i,g}$ are the coefficients to be determined, $m$ indicates the number of exponentials.
The formal analysis is simpler on the  $\alpha_j$ graded basis, where we obtain a more compact expression on the right-hand side,
\begin{equation}\label{eq:CF_definition_alpha}
\begin{split}
    U_{m}^{[2s]}(h) &:= \prod_{i=1}^m \exp\left(\sum_{j=1}^s x_{i,j} \alpha_j\right) = \exp(\Omega^{[2s]}(h)) + O(h^{2s+1}).
\end{split}
\end{equation}

The parameters $x_{i,j}$ represent the coefficients to be determined. We want to obtain a time-antisymmetric expression~\cite[Eq. 33]{blanes2006fourth}, so we impose
\begin{equation}\label{eq:time_symmetry_parameter_constraint}
    x_{m+1-i,j} = (-1)^{j+1}x_{i,j}.
\end{equation}
As a consequence, for a given value of $m$, we have $\frac{sm}{2}$ parameters to replicate $\Omega^{[2s]}(h)$ up to order $O(h^{2s+1})$. This imposes a minimum $m$ for each order $s$.
We can find coefficients $x_{i,j}$ by expanding both sides in the Taylor series and solving the equations to reproduce the equality~\cref{eq:CF_definition_alpha} to order $O(h^{2s+1})$. More sophisticated procedures exist too, see~\cite{blanes2006fourth}.

Once we have found the parameters $x_{i,j}$, we map the result to the univariate basis $\{A^{(g)}(h)\}$ using $(T^{(s)})^{-1}$~\cite{blanes2006fourth}. The resulting operator $\tilde{U}_{m}^{[2s]}$ contains univariate integrals that are approximated using Gaussian quadrature with coefficients $A_k$,
\begin{equation}\label{eq:CF_definition_quadrature}
\begin{split}
    \overline{U}_m^{[2s]}(h) &:=  \prod_{i=1}^m \exp\left(\sum_{k=1}^{s} z_{i,k} A_k h\right)= \exp(\overline{\Omega}^{[2s]}(h)) + O(h^{2s+1}).
\end{split}
\end{equation}
$U_m^{[2s]}$, $\tilde{U}_m^{[2s]}$ and $\overline{U}_m^{[2s]}$ will be referred to as \textit{quasi-Magnus operators}, as they approximate the evolution generated by the Magnus operators $\Omega^{[2s]}$, $\tilde{\Omega}^{[2s]}$ and $\overline{\Omega}^{[2s]}$ respectively, see~\cref{fig:Magnus_derivation}.

Finally each exponential is implemented on the quantum computer via a $2s$-order product formula $S_{2s}$, e.g. Trotter-Suzuki,
\begin{equation}\label{eq:CF_definition_trotter}
S_{2s}\left(\overline{U}_m^{[2s]}(h)\right):= \prod_{i=1}^m S_{2s}\left(\exp\left(\sum_{k=1}^{s} z_{i,k} A_k h\right)\right).
\end{equation}
This last step assumes we decompose exponents $\sum_{k=1}^{s} z_{i,k} A_k h$ into fast-forwardable terms. Throughout this paper, we say a Hamiltonian is \textit{fast-forwardable} to indicate the existence of quantum circuits that simulate the evolution it generates at a polylogarithmic cost in the evolution time and target accuracy. As such, it is close to what is sometimes called `exactly integrable' in the geometric integration literature~\cite{blanes2017concise}. Our definition is slightly more stringent than others in the quantum algorithmic literature, which only require sublinear cost in the evolution time~\cite{gu2021fast,fang2024time}.

The procedure explained in this section corresponds to the lower half of~\cref{fig:Magnus_derivation}, and steps~\ref{alg:step_CFQM_alphas} to~\ref{alg:step_CFQM_Trotter} in~\cref{alg:Steps_CFQMs}. Note that a variation of this method, referred to in this work by `split-operator', avoids the Trotter step~\ref{alg:step_CFQM_Trotter}, meaning the quadrature-based product formula in its original form can be directly implemented on the quantum computer without further approximation; however, this requires the Hamiltonian to be split to self-commuting fast-forwardable terms.
An important application would be electronic structure Hamiltonians where both the kinetic $T(t) \equiv T$ and Coulomb potential $V(t)$ are not only fast-forwardable but also self-commuting at different times. 
In such case, we would use a modification of the derivation above to find compositions of the form
\begin{align}
    W^{[2s]}_{m_s} = \prod_{j=1}^{m_s}\left[ e^{-i \theta_j T h} e^{-i \sum_k \nu_{j,k} V(t_k) h}\right].
\end{align}
We refer to~\cref{app:split_operator} for a detailed explanation on how to adapt the method presented above to find them.

\begin{algorithm}[H]
\caption{Steps to build and implement CFQMs. The three error terms are those comprising the definition of the CFQM (related to steps~\ref{alg:step_Magnus_Taylor_trunc} to~\ref{alg:step_CFQM_integrals}), the quadrature error (\ref{alg:step_CFQM_quadrature}), and product formula error (\ref{alg:step_CFQM_Trotter}). Through this procedure, $\{\alpha_j\}$ and $\{A^{(g)}\}$ will be treated symbolically; only the $\{A_k\}$ will be evaluated numerically.
\label{alg:Steps_CFQMs}}
\begin{algorithmic}[1]
        \State Substitute the symbolic Taylor expression of the Hamiltonian, see~\cref{eq:Taylor}, in the Magnus operator. Integrate the expression, obtaining the exact expression of the Magnus operator as a function of $\{\alpha_j\}$.\label{alg:step_Magnus_Taylor}
        \State Retain only the terms in the above expression that contain $\alpha_j$'s with $j = 1,\ldots,s$. These form the representation of the Magnus operator in a different Lie algebra. As a result we obtain $\Omega^{[2s]}$. For example, up to $s=3$ we obtain~\cref{eq:Omega2_1generators,eq:Omega4_2generators,eq:Omega6_3generators}. \label{alg:step_Magnus_Taylor_trunc}
        \State Use~\cref{eq:CF_definition_alpha,eq:time_symmetry_parameter_constraint} to compute the coefficients $x_{g,j}$ that define the CFQM $U^{[2s]}_m$ in the basis generated by $\{\alpha_j\}$ with $j=1,\ldots,s$. \label{alg:step_CFQM_alphas}
        \State Use $(T^{(s)})^{-1}$ to map $U^{[2s]}_m$ to the original Lie algebra generated by the univariate integrals $A^{(g)}(h)$, obtaining $\tilde{U}^{[2s]}_m$. \label{alg:step_CFQM_integrals}
        \State Use quadrature, e.g. Gaussian, to approximate each univariate integral $A^{(g)}$ in each exponential as a function of $\{A_k\}$, obtaining $\overline{U}^{[2s]}_m$. \label{alg:step_CFQM_quadrature}
        \State Use a product formula, e.g. Trotter-Suzuki, to approximate each exponential to order $O(h^{2s+1})$. We obtain $S_{2s}(\overline{U}^{[2s]}_m)$. \label{alg:step_CFQM_Trotter}
\end{algorithmic}
\end{algorithm}

\section{Error bounds}\label{sec:error}

To bound the overall error of implementing CFQMs, we will use the following triangle inequality, reflecting the steps in the derivation and implementation illustrated in \cref{fig:Magnus_derivation}, where we leave implicit the variable $h$:
\begin{equation}\label{eq:general_error_bound_CFMagnus}
\begin{split}
    &\left\|\exp(\Omega) -  S_{2s}\left(\overline{U}_m^{[2s]}\right) \right\| \leq \left\|\exp(\Omega) -\tilde{U}_{m}^{[2s]} \right\| + \left\| \tilde{U}_{m}^{[2s]} -\overline{U}_m^{[2s]} \right\| + \left\| \overline{U}_m^{[2s]} - S_{2s}\left(\overline{U}_m^{[2s]}\right) \right\|.
\end{split}
\end{equation}
The first term on the right-hand side of the inequality represents the $y_{i,g}$ parameter-fixing in \cref{eq:CF_definition_integrals}. Thus, we will refer to it as the error from the (univariate integral-based) CFQM definition. The second and third terms represent the quadrature and product formula errors. 

Bounding the first term represents the most important contribution of this work. The derivation procedure guarantees that the Taylor series of both terms in \cref{eq:CF_definition_integrals} match up to order $O(h^{2s+1})$~\cite{blanes2006fourth}. Thus, we decompose this error as
\begin{align}\label{eq:CFQM_definition_error}
    \left\|\exp(\Omega) - \tilde{U}_{m}^{[2s]}\right\|\leq \sum_{p=2s+1}^\infty\left( \left\|\exp(\Omega)_p\right\| +  \|\tilde{U}_{m,p}^{[2s]}\|\right),
\end{align}
where $\left\|\exp(\Omega)_p\right\|$ and $\|\tilde{U}_{m,p}^{[2s]}\|$ represent the norm of the $\Theta(h^p)$ term in the Taylor expansion of $\exp(\Omega)$ and $\tilde{U}_{m}^{[2s]}$, respectively. We will refer to them respectively as the Taylor series truncation error of the Magnus and quasi-Magnus operators.
Next, we present the main results and briefly comment on how to bound each error term.

\subsection{Taylor remainder of the Magnus operator}

The integer compositions $\mathcal{C}(p)$ of $p\in\mathbb{N}$ are a way of decomposing $p$ into sums of positive integers, where order matters. For instance, $\mathcal{C}(3) = \{[3],[2,1],[1,2],[1,1,1]\}$, where $[2,1]$ corresponds to $3 = 2+1$, and each element has a `dimension' equal to its size, e.g., $\dim ([1,2]) = 2$. We denote by $\mathcal{C}^z(p)$ the compositions of dimension $z$. Let us also uniformly bound the coefficients $a_j$ of the Taylor series of $A(t)$ by $c \ge \|a_{j}\|, \forall j$.

\begin{prop}\label{prop:expMagnus_Taylor}
Let $A(t)$ be a bounded operator, with its Taylor coefficients bounded by $c$. Let $h$ be the step size of the Magnus operator $\Omega$ and
$\exp(\Omega)_p$ be the $\Theta(h^p)$-term in the Taylor expansion of $\exp(\Omega)$. Then
%     &\sum_{p=2s+1}^\infty \|\exp(\Omega)_{p}\| \leq \sum_{p=2s+1}^\infty \left[\left(\frac{h}{2}\right)^p \cdot \right.\\
%      &\left.\sum_{\bm{k}\in\mathcal{C}(p)}\frac{1}{(\dim \bm{k})!} \prod_{l=1}^{\dim \bm{k}} \sum_{\bm{j}_l\in\mathcal{C}(k_l)} \frac{(2c)^{\dim \bm{j}_l}}{\dim \bm{j}_l}\prod_{\ell_l=1}^{\dim \bm{j}_l}  \frac{1}{j_{\ell_l}}\right].
\begin{equation}
\begin{split}
    &\sum_{p = 2s+1}^\infty\|\exp(\Omega)_{p}\| \leq  \sum_{p=2s+1}^\infty\left[\left(\frac{h}{2}\right)^{p} \sum_{\bm{k}\in\mathcal{C}(p)}\frac{1}{(\dim \bm{k})!} \prod_{l=1}^{\dim \bm{k}} \sum_{\bm{j}_l\in\mathcal{C}(k_l)} \frac{(2c)^{\dim \bm{j}_l}}{\dim \bm{j}_l}\prod_{\ell_l=1}^{\dim \bm{j}_l}  \frac{1}{j_{\ell_l}}\right].
\end{split}
\end{equation} 
\end{prop}

The proof can be found in \cref{sec:Error_Taylor_Magnus}. The key step in bounding the norm of these Taylor expansion terms is to start from the expression of the Magnus operator given in Refs.~\cite{mielnik1970combinatorial,moan2001convergence},
\begin{align}
    &\Omega(t_0, t_0+h) =\sum_{n=1}^\infty\int_{t_0}^{t_0+h} dt_n\ldots \int_{t_0}^{t_0+h} dt_1 L_n(t_n,\ldots, t_1) A(t_n)\ldots A(t_1),\nonumber
\end{align}
where
\begin{equation}
L_n(t_n,\ldots, t_1) = \frac{\Theta_n!(n-1-\Theta_n)!}{n!}(-1)^{n-1-\Theta_n},
\end{equation}
with $\Theta_n = \theta_{n-1,n-2} +\ldots +\theta_{2,1}$ and $\theta_{b,a}$ being the step function that is equal to $1$ when $t_b>t_a$ and $0$ otherwise. Consequently,
\begin{equation}
    |L_n(t_n,\ldots, t_1)|\leq \frac{(n-1)!}{n!} = \frac{1}{n}.
\end{equation}
Then, we substitute the Taylor series of the Hamiltonian, and group terms by their $\Theta(h^p)$ order. This strategy allows us to use the submultiplicativity of the spectral norm, and generate triangle inequalities. Manipulating the expressions, we obtain the desired result.

\subsection{Taylor remainder of the quasi-Magnus operator}

Now we turn to the CFQM operator from \cref{eq:CF_definition_integrals}. We first need to define an extension of $x_{i,j}$ for values $j>s$,
\begin{equation}
    \overline{x}_{i,j} := \sum_{g=0}^{s-1} y_{i,g} \frac{1-(-1)^{g+j}}{(g+j)2^{g+j}}.
\end{equation}
This extension is needed as we want to account for all Taylor coefficients $a_j$ in~\cref{eq:Taylor}, not just up to $j = s$.
Then, we uniformly bound $\|\overline{x}_{i,j}\alpha_{j}\| = \|\overline{x}_{i,j}a_{j-1}\|h^{j} \leq \bar{c} h^{j}$ by the smallest possible $\bar{c}$. %Since we know the values of $x_{i,j}$ from the formal definition, this is a similar condition to bounding the coefficients of the Taylor expansion of Hamiltonian.

\begin{prop}\label{prop:CFQM_Taylor}
Let $h$ and $\bar{c}$ be defined as above. Let $\tilde{U}_{m}^{[2s]}$ be the $2s$-order $m$-exponential commutator-free quasi-Magnus operator, expressed in terms of the univariate integrals $A^{(g)}(h)$ in \cref{eq:CF_definition_integrals}.
Let $\tilde{U}_{m,p}^{[2s]}$ be the $\Theta(h^p)$-term in its Taylor expansion.
Then
    \begin{equation}
\sum_{p=2s+1}\|\tilde{U}_{m,p}^{[2s]}\| \leq \sum_{\substack{p=\\2s+1}}^\infty h^{p} \sum_{
        \substack{
            \bm{w}\in\mathcal{C}(p)
            }
    }
    \sum_{\substack{\bm{k} : \sum_{i=1}^m k_i\\ = \dim\bm{w}}}
    \frac{\bar{c}^{k_1}\cdots \bar{c}^{k_m}}{k_1!\cdots k_m!}.
\end{equation}
\end{prop}
The proof, in \cref{sec:Error_Taylor_quasiMagnus}, proceeds similarly to the previous result. %In the last expression note that the summation is over the \textit{weak} compositions of $\dim\bm{w}$ of size $m$.%, though we have chosen to not introduce further notation for this.

\subsection{Gauss-Legendre quadrature error}
Next, we estimate the error incurred by approximating the $A^{(g)}(h)$ integrals in the CFQM with a Gauss-Legendre quadrature, with proof in \cref{sec:Quadrature_error}.
\begin{prop}\label{prop:Quadrature}
Let $\tilde{U}_{m}^{[2s]}$ represent the commutator-free Magnus operator, as a function of the univariate integrals and with coefficients $y^{(i)}_j$, see \cref{eq:CF_definition_integrals}. Let $\overline{U}_m^{[2s]}$ be the operator where the integrals have been approximated via a Gauss-Legendre quadrature. Let $R_{s}(A)$ denote the residual from approximating the integral $A$ with quadrature of order $2s$.
Then,
\begin{align}\label{eq:quadrature_error}
   \left\| \tilde{U}_{m}^{[2s]} -\overline{U}_m^{[2s]} \right\| &\leq \sum_{i=1}^m\sum_{g=0}^{s-1} \left\|y_{i,g} R_{s}(A^{(g)}(h))\right\|\\
    &\leq \frac{h^{2s+1}(s!)^4}{(2s+1)((2s)!)^3}\sum_{i=1}^m\sum_{g=0}^{s-1}
   \left( \frac{|y_{i,g}|}{h^g} \sum_{j = 0}^\infty \frac{(j+2s)!}{j!}\frac{h^j}{2^j}\|a_{j+2n-g}\|\right).
\end{align}
\end{prop}
To compute this error term, we apply the following well-known result for unitary operators 
\begin{align}\label{eq:error_exponentials}
   \|e^{X_1}\cdots e^{X_m} - e^{Y_1}\cdots e^{Y_m}\|&\leq \sum_{j=1}^m \|e^{X_j}-e^{Y_j} \|,
\end{align}
along with the following:
\begin{cor}[Corollary A.5 in \cite{childs2021theory}]\label{cor:Distance_bound}
Given $\mathcal{H}$ and $\mathcal{G}$ two antihermitian continuous operator-valued functions defined on $\mathbb{R}$,
\begin{equation}
\begin{split}
\left\|\mathcal{T}\exp\left(\int_{t_1}^{t_2}d\tau \mathcal{H}(\tau)\right)-\mathcal{T}\exp\left(\int_{t_1}^{t_2}d\tau \mathcal{G}(\tau)\right)\right\|\leq \int_{t_1}^{t_2}d\tau \|\mathcal{H}(\tau)-\mathcal{G}(\tau)\|.
\end{split}
\end{equation}
\end{cor}
%Consequently, by triangle inequality,
%\begin{align}
%    \|e^{X_1}\cdots e^{X_m} - e^{Y_1}\cdots e^{Y_m}\| \leq \sum_{j=1}^m \|X_j-Y_j\|.
%\end{align}
We then compute the error from using quadrature to approximate each exponential $X_i = \sum_{g=0}^{s-1}y_{i,g} A^{(g)}(h)$ in $\tilde{U}_{m}^{[2s]}$. The error from approximating the univariate integral $A^{(g)}(h)$ with quadrature
\begin{equation}
    \int_a^b f(x) dx = \sum_{i=1}^{n}w_if(c_i)+ R_n
\end{equation}
can be bounded using the expression~\cite[p. 146, Chapter 5]{kahaner1989numerical}
\begin{equation}
    R_n\left(\int_a^b f\right) = \frac{(b-a)^{2n+1}(n!)^4}{(2n+1)((2n)!)^3}f^{(2n)}(\xi),\qquad a<\xi<b.
\end{equation}
These techniques suffice to compute $\left\|y_{i,g} R_{s}(A^{(g)}(h))\right\|$ and prove the statement.

A key limitation of bound~\cref{eq:quadrature_error} is its convergence as $O(h^{s+2})$ -- instead of the desirable $O(h^{2s+1})$ proven in~\cite{blanes2020convergence} -- due to the presence of a factor of $h^{g}$ in the denominator. We will see in our numerical experiments,~\cref{fig:error_weights}, that this error contribution is negligible. Yet, if we want to ensure the $O(h^{2s+1})$ convergence, we could do so by increasing the number of quadrature points of the integrals to $\lceil \frac{3s}{2}\rceil$. This would not increase the number of exponentials to approximate via Trotter, but would modify the time-independent Hamiltonians in the exponents of~\cref{eq:CF_definition_trotter}, $\sum_{k}z_{i,k}A_kh$.

\subsection{Product formula error}
The final error term is the product formula error $\|\overline{U}_m^{[2s]}-S_{2s}(\overline{U}_m^{[2s]})\|$, or in other words, the error due to splitting the exponentials in $\overline{U}_m^{[2s]}$ into fast-forwardable exponentials. The implementation cost of such exponentials grows at most polylogarithmically in the simulation time, and optimized methods for their implementation are known~\cite{motzoi2017linear}.
We refer to error bounds derived in the literature for Trotter-Suzuki product formulas, which can be directly applied to each operator $\exp\left(\sum_{k}z_{i,k}A_kh\right)$ in $\overline{U}_m^{[2s]}$~\cite{childs2021theory}; and then combined with~\cref{eq:error_exponentials} for a full account of the errors incurred during the usage of CFQMs.

\section{Numerical results}\label{sec:results}

% In contrast to other techniques, there is no simple closed formula.
In contrast to other methods, CFQM errors do not have a simple expression that allows us to straightforwardly understand the cost dependence on key variables. However, we can numerically evaluate their performance. We analyze the time-dependent Heisenberg model
\begin{equation}\label{eq:Heisenberg_Hamiltonian}
    -iA(t) = \frac{1}{4n}\sum_{i=1}^{n-1} \vec{\sigma}_i\vec{\sigma}_{i+1} + \frac{1}{4n}\sum_{i=1}^{n} \cos(\phi_i + \omega_i t)\sigma_i^z.
\end{equation}
We have added the factor $1/4n$ to ensure $\|A(t)\| \le 1$ and simplify the analysis.  We assume that $\{\omega_i,\phi_i\}$ are arbitrarily chosen such that 
\begin{equation}
    \frac{1}{j!}\left\|\frac{d^j A(t)}{dt^j}\right\|\leq c\leq 1, \quad \forall j\in\mathbb{N}.
\end{equation}
\begin{figure*}[t]
    \centering
    \includegraphics[width=0.373\textwidth]{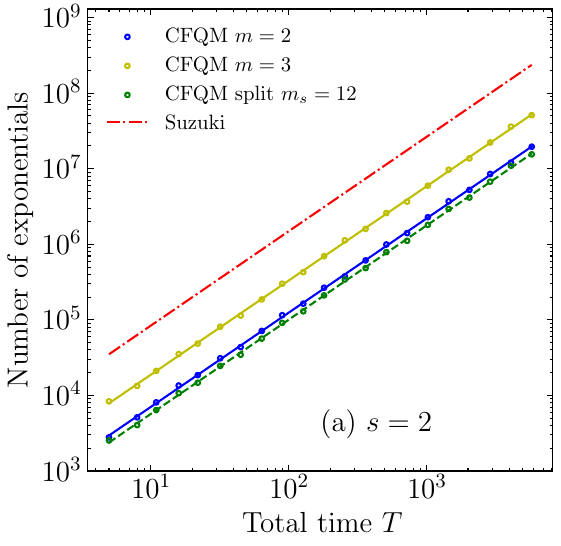}
    \includegraphics[width=0.35\textwidth]{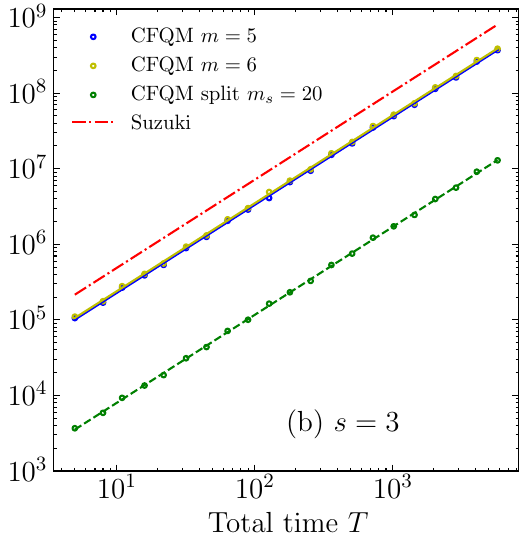}
    \caption{Cost scaling of several CFQMs (including split-operator ones from~\cref{app:split_operator}) and the Suzuki operator for a normalized Heisenberg Hamiltonian~\cref{eq:Heisenberg_Hamiltonian} as a function of the total evolution time $T$. The number of spins is fixed at $n = 128$ and the total target error is $\epsilon = 10^{-3}$, similar to that used in~\cite{childs2018toward,childs2019nearly,childs2021theory}. The $y$ axis indicates the number of fast-forwardable exponentials to be implemented, and in all cases it grows as $T^{1+1/2s}$. (a) 4th-order operators: CFQMs save up to an order of magnitude of cost compared with the Suzuki method. (b) 6th-order operators. The 6th-order Suzuki method displays a cost just twice as large as the two 6th-order non-split CFQM bounds for $n = 128$, due to the dominant Trotter error contribution. Since the known Trotter bounds scale better than Suzuki and Taylor errors with $n$, once the latter dominates, all analyzed CFQMs will exhibit better cost than the Suzuki method of the same order, see~\cref{fig:time_n_scaling,fig:error_weights}.}
    \label{fig:scaling_time}
\end{figure*}

We analyze the cost of four CFQMs of order 4 (with $m=2$ or $m=3$ exponentials) and order 6 (with $m=5$ and $m=6$ exponentials) from Ref.~\cite{blanes2006fourth}; as well as the cost of two `split-operator' CFQMs that avoid the last Trotter decomposition step, as we review in~\cref{app:split_operator}. These split-operator CFQMs can be found in tables 2 and 3 of Ref.~\cite{blanes2006splitting}, under the names GS$_{6}$-$4$ and GS$_{10}$-$6$. The requirement to use them is that the Hamiltonian can be split into two self-commuting and fast-forwardable Hermitian terms. As we will see, these requirements apply to~\cref{eq:Heisenberg_Hamiltonian}.

\subsection{Time-dependent Heisenberg Hamiltonians}
 
To compute the simulation error, we apply a Trotter-Suzuki product formula decomposition to each exponential in $\overline{U}_m^{[2s]}$, see in~\cref{eq:CF_definition_trotter}. We compute the Trotter error of such approximation with the results in~\cite{childs2019nearly}. To do so, one starts by splitting each exponent $\sum_{k=1}^s z_{i,k} A_k h$ into two fast-forwardable terms $-iBh$ and $-iCh$, e.g., terms whose Hamiltonian simulation cost depends only logarithmically on the simulation time. For each $A_k$, we select%~\cite[Eq. 51 appendix]{childs2019nearly}
%\begin{align}
%    A = H_{\text{odd}}:= \sum_{k=1}^{n/2}H_{2k-1,2k},\\
%    B = H_{\text{even}}:= \sum_{k=1}^{n/2-1}H_{2k,2k+1},
%\end{align}
~\cite[Eqs 123,124]{childs2021theory}
\begin{align}\label{eq:odd_Heisenberg_external}
    &H_{k,\text{odd}}= \frac{1}{4n}\sum_{l=1}^{\left\lfloor \frac{n}{2}\right\rfloor} \left(\vec{\sigma}_{2l-1}\vec{\sigma}_{2l} + \cos(\phi_{2l-1} + \omega_{2l-1} t_k)\sigma_{2l-1}^z\right), \\
    \label{eq:even_Heisenberg_external}
    &H_{k,\text{even}} = \frac{1}{4n}\sum_{l=1}^{\left\lceil \frac{n}{2}\right\rceil -1} \left(\vec{\sigma}_{2l}\vec{\sigma}_{2l+1}+  \cos(\phi_{2l} + \omega_{2l} t_k)\sigma_{2l}^z\right),
\end{align}
with $t_k = t_0 + c_k\frac{h}{2}$. Thus, for each exponential index by $i$, $B :=\sum_{k=1}^s z_{i,k} H_{k,\text{odd}}$ and $C :=\sum_{k=1}^s z_{i,k} H_{k,\text{even}}$.
Then, we define:
\begin{dfn}[Canonical product formula~\cite{childs2019nearly}]
    Let $H = B+C$ be a time-independent Hamiltonian, with $B$ and $C$ Hermitian operators. We say $S$ is a canonical product formula when it has the form
    \begin{equation}
        S(t):= \left(e^{-it\xi_{\varsigma}C}e^{-it\beta_{\varsigma}B}\right)\ldots \left(e^{-it\xi_1C}e^{-it\beta_1B}\right).
    \end{equation}
The parameter $\varsigma$ indicates the number of stages. Let $u$ be an upper bound to the coefficients
\begin{equation}
    \max\{|\xi_1|,\ldots,|\xi_{\varsigma}|,|\beta_1|,\ldots,|\beta_{\varsigma}|\}\leq u.
\end{equation}
We say $S$ has order $p$ if $S(t) = e^{-iHt} + O(t^{p+1})$, and we will denote it as $S_p(t)$. We shall use the notation $(\varsigma,p,u)$-product formula to denote this canonical product formula. The $2s$-th order Trotter-Suzuki formula $S_{2s}(t)$ is an $(\varsigma,p,u)$-formula with $\varsigma = 2\cdot 5^{s-1}$, $p=2s$ and $u=1$.
\end{dfn}
The error of the Trotter formula for the unnormalized Heisenberg model scales as~\cite[Eqs. (47-58) in Appendix]{childs2019nearly}
\begin{equation}\label{eq:Trotter_error}
\begin{split}
    \|S_p(t)-e^{-iHt}\|&\leq\sum_{k=1}^\varsigma n(2k-1)^p(ku)(2u)^p(2k-2)^p\frac{t^{p+1}}{(p+1)!}\\
    &+ \sum_{k=1}^\varsigma n(2k+1)^p(ku)(2u)^p(2k)^p\frac{t^{p+1}}{(p+1)!}.
\end{split}
\end{equation}
Since we are implementing a normalized Heisenberg model, we absorb the factor of $(4n)^{-1}$ in $t$. We similarly include in $t$ the prefactor $\sum_{k=1}^s |z_{i,k}|$ for each of the $i=1,\ldots,m$ exponentials. Thus, calling $Z_i = \frac{\sum_{k=1}^s |z_{i,k}|}{4n}$ and substituting the remaining parameters of the $2s$-th order Trotter-Suzuki formula, we get
\begin{equation}\label{eq:Trotter_error_adapted}
\begin{split}
    \left\|S_{2s}(h)-\exp\left(\sum_{k=1}^s z_{i,k} A_k h\right)\right\|&\leq\sum_{k=1}^{2\cdot 5^{s-1} }n(2k-1)^{2s}k(4k-4)^{2s}\frac{(Z_ih)^{2s+1}}{(2s+1)!}\\
    &+ \sum_{k=1}^{2\cdot 5^{s-1}} n(2k+1)^{2s}k(4k)^{2s}\frac{(Z_ih)^{2s+1}}{(2s+1)!}.
\end{split}
\end{equation}
This equation, in combination with \cref{eq:error_exponentials}, suffices to compute the Trotter product formula error. In the following section, we will use all the bounds we discussed to numerically estimate the cost of implementing CFQMs for the Heisenberg Hamiltonian in~\cref{eq:Heisenberg_Hamiltonian}.

\subsection{Comparison with the Suzuki method}

There are three well-known broad techniques in quantum algorithms to simulate time-dependent Hamiltonians: the Suzuki method~\cite{wiebe2010higher}, the continuous qDRIFT technique~\cite{poulin2011quantum,berry2020time}, and the Dyson series method~\cite{kieferova2019simulating,low2018hamiltonian,berry2020time}. The Suzuki method is an adaptation of Trotter-Suzuki product formulas for time-dependent Hamiltonians~\cite{suzuki1990fractal,suzuki1991general,suzuki1993general}. The continuous qDRIFT is the generalization of the time-independent qDRIFT technique~\cite{campbell2019random}, whose main disadvantage is the quadratic scaling with respect to the simulation time. Finally, the Dyson series method is a linear combination of unitaries approach that shares significant similarities with the Taylor series Hamiltonian simulation method, and similarly requires employing oblivious amplitude amplification~\cite{berry2015simulating}. Finally, there is always the option of approximating the time evolution using the Hamiltonian evaluated at the midpoint of each segment, which is called the exponential midpoint method and corresponds to the second-order Magnus operator~\cite{gomez2018propagators}. We review and compare these techniques with CFQMs in~\cref{app:Comparison}.

The most natural comparison of CFQMs is with other product formulas and specifically with the Suzuki technique of Ref.~\cite{wiebe2010higher}. They rely on the following definition.
\begin{dfn}[$\Lambda$-$P$-smooth operators]
The set of operators $\{O_j\}$ is said to be
    $\Lambda$-$P$-smooth if for all $O_j$, all $\tau \in [t_0, t_0+h]$ and all $p\in\{1,\ldots,P\}$,
\begin{equation}
    \Lambda \geq \left(\sum_{j=1}^m \|\left.\partial_t^p O_j(t)\right|_\tau\|\right)^{1/(p+1)}.
\end{equation}
\end{dfn}

\begin{figure*}[t]
    \centering
    \includegraphics[width=0.373\textwidth]{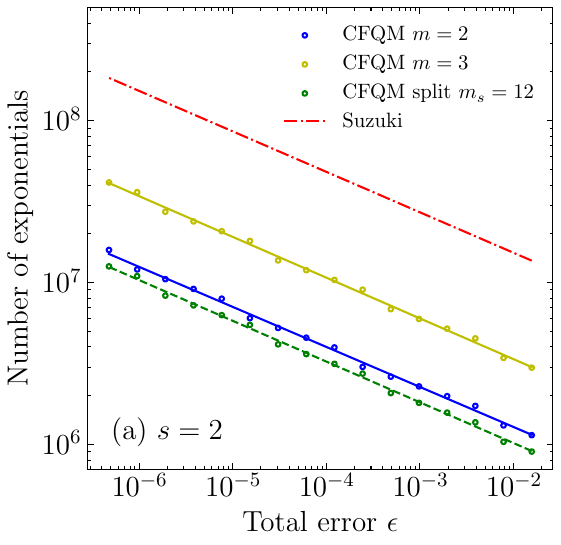}
    \includegraphics[width=0.35\textwidth]{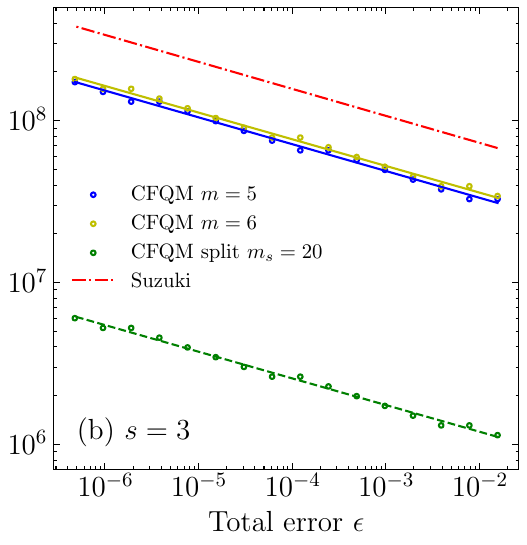}
    \caption{Cost scaling of several CFQMs (including split-operator ones from~\cref{app:split_operator}) and the Suzuki operator for a normalized Heisenberg Hamiltonian~\cref{eq:Heisenberg_Hamiltonian} as a function of the target error $\epsilon$. Plots are given for (a) 4th-order operators, and (b) 6th-order operators. The number of spins is fixed at $n = 128$ and the total evolution time at $T = 1024$. The $y$ axis indicates the number of fast-forwardable exponentials to be implemented, and in all cases, they grow as $\epsilon^{-1/2s}$.}
    \label{fig:scaling_error}
\end{figure*}

The cost of implementing one segment of length $h$ using the Suzuki method is measured via the number $N$ of fast-forwardable exponentials as a proxy, which is upper bounded as \cite[Theorem 1] {wiebe2010higher}:
% \begin{thm}[Theorem 1 in~\cite{wiebe2010higher}]
%     Let the Hamiltonian $A(t)$ be decomposed in $q$ $\Lambda$-$2s$-smooth Hermitian operators whose exponential can be fast-forwarded. Let also $\epsilon\leq \frac{9}{10}\left(\frac{5}{3}\right)^s \Lambda h$. Then we can construct an approximation $\Tilde{U}$ to the true time-dependent evolution operator $U$, with error at most $\epsilon$ and number of fast-forwardable exponentials given by
\begin{equation}\label{eq:Suzuki_cost}
    N\leq \left\lceil 
    3q\Lambda h s\left(\frac{25}{3}\right)^s\left(\frac{\Lambda h}{\epsilon}\right)^{\frac{1}{2s}}
    \right\rceil,
\end{equation}
% \end{thm}
where we have assumed a Hamiltonian $A(t)$ that can be decomposed to $q$ many $\Lambda$-$2s$-smooth Hermitian fast-forwardable operators, and the target error satisfies $\epsilon\leq \frac{9}{10}\left(\frac{5}{3}\right)^s \Lambda h$. We choose $\Lambda = 1 = c$ for the comparisons, as both represent an upper bound to the norm of the terms in the Hamiltonian and their derivatives.

We now compare the Suzuki algorithm to ours by using three key variables, the total simulation time $T$, the target error $\epsilon$, and the number of spins $n$. 
The code used to compute the numerical results of this section may be found in~\url{https://github.com/XanaduAI/CFQMagnus}.
The first comparison is indicated in~\cref{fig:scaling_time}, where we compute the cost implied by CFQM bounds and~\cref{eq:Suzuki_cost} when simulating the time-dependent Heisenberg model of~\cref{eq:Heisenberg_Hamiltonian} for $n = 128$ and $\epsilon = 10^{-3}$. There, we observe that both the Suzuki and CFQM operators scale asymptotically as $O(T^{1+1/2s})$, with CFQMs displaying a lower constant factor overhead, saving up to an order of magnitude in subfigure (a) for 4th-order methods.
% Similarly favorable behavior of CFQMs may be found in subfigure (b), which evaluates 6th-order methods.
Split-operator CFQMs perform particularly well in both cases, with a very slight edge for the 6th-order due to its better scaling. 
% Their contrast to the Suzuki method highlights the potential for CFQMs to become the new state-of-the-art to simulate time-dependent spin systems in early fault-tolerant quantum computers, though further research is required to confirm this theoretically and numerically. Non-split 4th-order CFQMs are also more cost-efficient than 4th-order Suzuki, as can be seen in~\cref{fig:scaling_time} (a). 

In contrast,~\cref{fig:scaling_time} (b) shows the Suzuki method complexity closer to that of the two 6th-order non-split CFQMs. This is due to the Trotter-Suzuki error in~\cref{eq:Trotter_error_adapted}, which dominates in that $\epsilon$ and $n$ regime, see~\cref{fig:time_n_scaling,fig:error_weights}.
% Known Trotter error bounds scale better with the number of spins $n$ than the Taylor error bounds of CFQMs, or that of the Suzuki method; but here they display a larger prefactor. 
% Specifically, 
Trotter methods display a cost almost linear in $n$ lattice Hamiltonians~\cite{childs2019nearly}, in contrast to Suzuki's and our CFQM bounds.
%It should be noted that Trotter methods display a cost almost linear in $n$~\cite{childs2019nearly}, while the available bounds for the Suzuki method and the Taylor bounds of CFQMs imply a superquadratic cost on the number of spins of an unnormalized Heisenberg Hamiltonian: one factor of $n$ due to the larger norm, and another $\tilde{O}(n)$ due to the cost of implementing each exponential over $n$ spins. 
This can be seen more clearly in~\cref{fig:time_n_scaling}, where we choose $n = T$, as done in several literature references, see~\cite{haah2021quantum}. This same figure can be interpreted as the scaling of our bounds at fixed target accuracy and simulation time for unnormalized Heisenberg Hamiltonians (e.g.~\cref{eq:Heisenberg_Hamiltonian} without the $1/n$ factor) of increasing size $n$. Non-split CFQMs exhibit two regimes, first one where Trotter error dominates, and another where Taylor errors become more important, see~\cref{fig:error_weights}. 
% In the former, we can observe a very flat cost growth, consistent with the almost linear cost of Trotter methods once we account for the fact that implementing each exponential requires a cost $\tilde{O}(n)$. Thus, this result raises the question of whether tighter error bounds displaying quasi-linear behavior on $n$ exist for the CFQMs and the Suzuki method. 

Finally, one may wonder about the scaling of the cost of CFQMs with respect to the target error. In~\cref{fig:scaling_error}, we show that CFQMs' asymptotic behavior is $O(\epsilon^{-1/2s})$ at fixed $T = 1024$ and $n=128$. This scaling is again the same as prescribed by~\cref{eq:Suzuki_cost}. %Overall, we find that the 6th-order split-operator CFQM GS$_{10}$-$6$ is particularly efficient, and CFQMs display again a better constant factor overhead than the Suzuki method. 
%If split-operator CFQMs are not an option, then the second Magnus order operator, e.g. the exponential midpoint method, is the best alternative until $T = n \approx 100$ at $\epsilon \leq 10^{-3}$. This is a special case of the Magnus operator, as many of the intermediate transformations are not necessary to compute it. Consequently, only the $\exp(\Omega)$ Taylor truncation error and the Trotter error contribute to its error bound. 
%If one is interested in simulating larger systems or with more stringent precision requirements, the 4th-order CFQM becomes the most inexpensive method until very high values of $\epsilon^{-1}$ and $T = n$. Another key conclusion is that for large $n$ (e.g. when the Trotter error no longer dominates), no Suzuki method of any order has been found to be more cost-efficient than the CFQM of the same order and lowest number of exponentials.
Additionally, in~\cref{fig:step_error_cost} of the supplementary material, we indicate the error incurred per step of size $h$ for each CFQM. Note that the $s = 1$ CFQM corresponds exactly to truncating the Magnus expansion at its first term~\cref{eq:Omega_1}, and then approximating the integral by its midpoint value. As such, the CFQM Taylor error from~\cref{sec:Error_Taylor_quasiMagnus} does not contribute to its overall cost. This technique is also called the \textit{exponential midpoint rule}~\cite{gomez2018propagators}.

Apart from reducing $h$ or increasing the order $s$ of the CFQM, there is a third option to reduce the simulation error: using multi-product formulas~\cite{childs2012LCUs,low2019well}. They have most of the advantages of product formulas, but they only incur a polylogarithmic cost in the precision, similar to LCU methods. Multi-product formulas were originally proposed in the classical computing geometrical integration literature ~\cite{blanes1999extrapolation,blanes2005raising,chin2010multi}, and can be applied to both the Suzuki method and CFQMs, as we show in theory in~\cref{app:Multi-product formulas}~\cite{geiser2011multi,low2019well}.

%More generally, we believe CFQMs will show an advantage in those regimes where Trotter is more appropriate than LCU techniques.

% However, we can numerically estimate the cost.

% We observe two regimes where different error dominate.

% 

\section{Conclusion}

% What we have done
In this paper, we have analyzed commutator-free quasi-Magnus operators (CFQMs) as a technique to implement time evolution under time-dependent Hamiltonians. The key contribution of this paper is providing a bound to the error of this method. This estimation allows us to find regimes where these operators outperform the alternative product formula techniques. Overall, we have found the split-operator CFQMs to exhibit a particularly performant behavior.

% Future work:
% 1. Arbitrary order formulas
% 2. Improving bounds
% 3. Extension to non-unitary dynamics.
From a theoretical perspective, it would be desirable to find methods to systematically compute CFQM coefficients to arbitrary order~\cite{blanes2006fourth,alvermann2011high,blanes2017high,blanes2018time}. As far as we know, our work is the first attempt at finding the error bounds, which include error expressions that are quite different from those of other time-dependent simulation methods. Yet, as noted above our bound for the quadrature error is disappointing because it does not achieve $O(h^{2s+1})$ convergence we know CFQMs achieve~\cite{blanes2020convergence}.
Further, it is likely that our error bounds are not tight, and further work may be able to improve them, either by deriving better bounds for each of the error terms, or via alternative constructions altogether. For instance, the error bounds we derive exhibit worse scaling than Trotter's quasi-linear behavior in $n$ for the Heisenberg Hamiltonian in~\cref {eq:Heisenberg_Hamiltonian}~\cite{childs2019nearly}. It would be desirable to find out whether it is possible to find alternative $O(n^{1+o(1)})$ bounds, or if it is possible to achieve commutator scaling and weak dependence on the derivatives shown in LCU-based Magnus techniques~\cite{fang2024time}. Explicit error bounds should be considered for cases where the Hamiltonian is unbounded too~\cite{hochbruck2003magnus,singh2018high,blanes2018convergence,blanes2020convergence,an2021time}, as well as for other CFQM constructions~\cite{alvermann2011high,blanes2017high,blanes2018time}. The advanced techniques reported in~\cite{blanes2018convergence,blanes2020convergence} might be a good starting point for that endeavor.

A very useful next step would be the development of a CFQM operator tailored to periodic Hamiltonians, the counterpart to the Floquet evolution algorithm developed in the LCU paradigm~\cite{mizuta2023optimal}. While there exists a Floquet-Magnus operator, see~\cite{casas2001floquetmagnus}, its commutator-free approximations have not yet been developed.

Another field potentially deserving more attention is the application of CFQMs to quantum algorithms for the resolution of differential equations. The Magnus technique has been applied to study open systems dynamics~\cite{blanes2017high}, and generally integrate differential equations of the form $x' = A(t) x $~\cite{blanes2006fourth}. This is in contrast to most of the quantum algorithm literature, which leverages linear algebra and finite method techniques~\cite{berry2014high,berry2017quantum,arrazola2019quantum,childs2020quantum}. However, since these dynamics may not be unitary, the techniques presented here will need to be developed further~\cite{cao2023quantum}. Additionally, CFQMs could become useful to implement interaction-picture algorithms that reduce the cost of the simulation when the Hamiltonian can be split into two components with different energy scales~\cite{low2018hamiltonian,bosse2024efficient, sharma2024hamiltonian}.

% Importance of time dependent Hamiltonian simulation for ISQ: dynamics
The area within quantum computing where these methods may be particularly useful is in the development of early fault-tolerant quantum algorithms, in particular for simulating spin models~\cite{park2023hardness,childs2018toward}.
As we have seen in~\cref{sec:results}, CFQMs, and particularly split-operator versions, seem particularly well suited to simulate time-dependent counterparts of spin systems.
% Product formulas are particularly well suited for this regime~\cite{childs2021theory}, and CFQM operators are one instance of product formulas.
% Further, quantum dynamics problems are also considered to be less resource-intensive than ground-state problems. Such dynamics problems often involve interactions with external fields, or equivalently, evolution under time-dependent Hamiltonians. 
As such, we believe they could become an important Hamiltonian simulation tool in the near future.

\section{Acknowledgements}

We thank Sergio Blanes, Fernando Casas and Jose Ros for answering questions about the details of commutator-free quasi-Magnus operators.% We also thank Utkarsh Azad for improving the code quality.

\clearpage

\appendix
%\onecolumngrid

\section{Comparison with time-dependent simulation algorithms~\label{app:Comparison}}

In this appendix, we provide further explanations on some of the most prominent algorithms to implement time-dependent Hamiltonian simulation; and compare their cost with CFQMs. Other methods exist in the literature, but they either have stronger assumptions such as periodicity \cite{mizuta2023optimal}, are variations of the ones presented here \cite{chen2021quantum,an2021time,an2022time,fang2024time}, or do not have cost estimates yet \cite{cao2023quantum}.

\begin{figure*}[t]
    \centering
\includegraphics[width=0.373\textwidth]{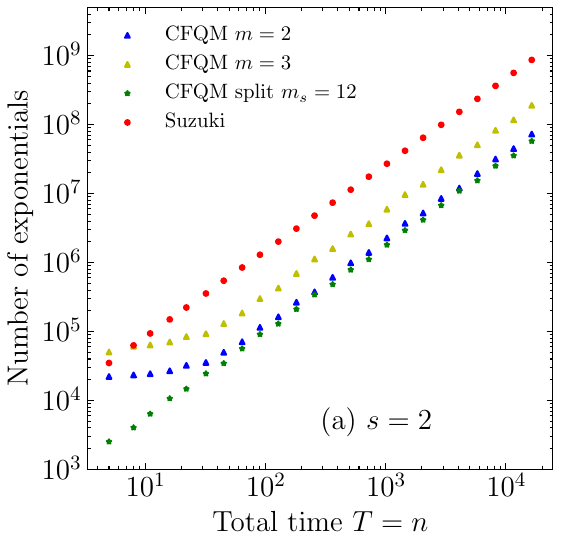}
\includegraphics[width=0.35\textwidth]{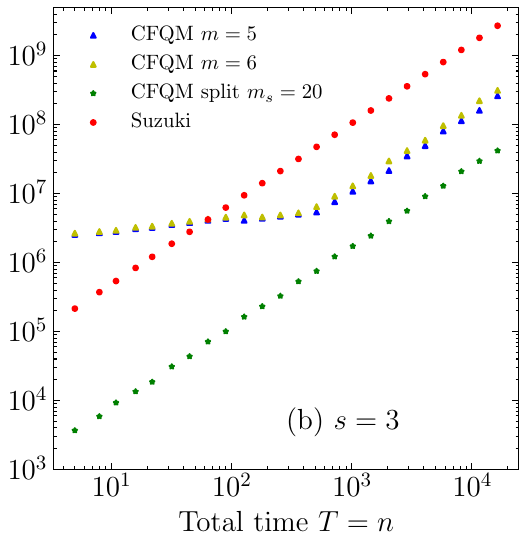}

    \caption{Cost of simulating the time-dependent Heisenberg Hamiltonian from~\cref{eq:Heisenberg_Hamiltonian} with a variety of non-split CFQMs $S(\overline{U}_m^{[2s]})$; split CFQMs $\overline{W}^{[2s]}_{m_s}$, see~\cref{app:split_operator}; and the Suzuki method~\cref{eq:Suzuki_cost}, for time $T = n$ and target error $\epsilon = 10^{-3}$. An alternative but equivalent picture is that of an unnormalized Heisenberg Hamiltonian (eg,~\cref{eq:Heisenberg_Hamiltonian} without the $1/n$ factor) being simulated for a fixed amount of time $T = 1$ and accuracy $\epsilon = 10^{-3}$, with an increased number of spins $n$. The $y$ axis indicates the number of fast-forwardable exponentials to be implemented, i.e. exponentials that can be simulated for time $t$ with cost $O(\log t)$. The CFQMs were taken from the literature~\cite{blanes2006fourth,alvermann2011high,blanes2006splitting}. %Since the Trotter product formula error is dominant in many cases, the split-operator approaches that forego its usage achieve a lower cost than their general counterparts. 
    The two regimes we observe for the non-split CFQMs are due to the different scaling of the dominant errors. The Trotter error is dominant at first, whereas the Taylor errors dominate when $T$ is large, see~\cref{fig:error_weights}. When Taylor error dominates, we empirically observe that the cost of CFQMs scales as $O(T^{1+1/2s})$, similar to the Suzuki method.}
    \label{fig:time_n_scaling}
\end{figure*}

\begin{figure*}
    \centering
    \includegraphics[width=0.225\textwidth]{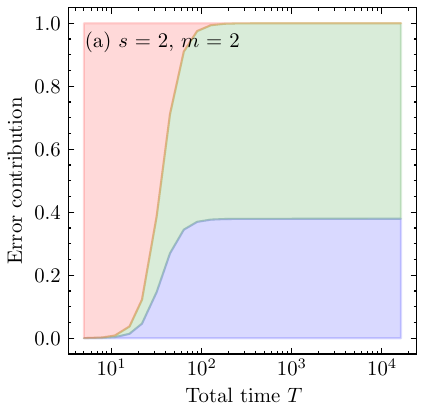}
\includegraphics[width=0.21\textwidth]{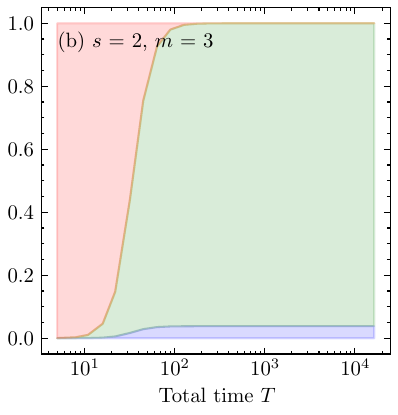}
\includegraphics[width=0.21\textwidth]{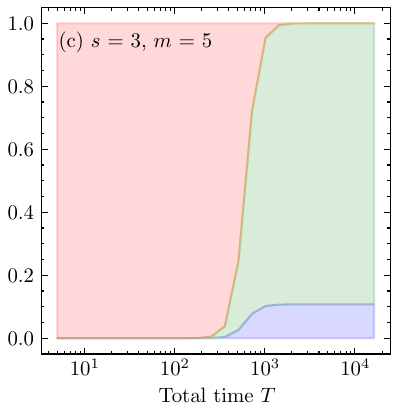}
\includegraphics[width=0.323\textwidth]{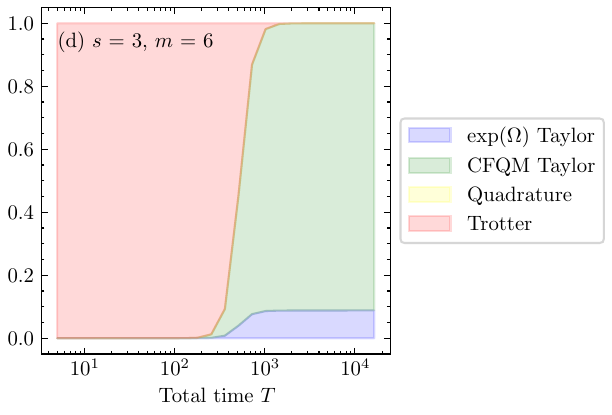}
\caption{Percentage of the contribution of each error source to the final target error in the non-split CFQMs of~\cref{fig:time_n_scaling}. We can observe the different behavior in the small and large $n$ corresponds to a change in the dominant error: from Trotter to Taylor error.\label{fig:error_weights}}
\end{figure*}

We also include numerical results highlighting the CFQMs cost scaling with respect to several variables, such as simulation time,~\cref{fig:scaling_time} or target error~\cref{fig:scaling_error}. When we select $T=n$, CFQMs that employ Trotter techniques exhibit the behavior corresponding to the sum of two power-laws~\cref{fig:time_n_scaling}. When $T$ is small, the Trotter error dominates. In contrast, for large $T$ the Taylor error contributions become more important, see~\cref{fig:error_weights}. In the latter regime, we empirically observe the number of exponentials to be implemented follows a power law behavior $O(T^{1+1/2s})$, similar, as we will see, to Suzuki's method~\cite{wiebe2010higher}. On the other hand, when Trotter error dominates the growth is much smaller, close to linearly in $n=T$ in the number of gates~\cite{childs2019nearly}. Since implementing each exponential will take $O(n)$ gates, the curve in the figure indicating the number of exponentials is almost flat.

\subsubsection{Exponential midpoint rule}

The simplest method to simulate time-dependent Hamiltonians is known as the exponential midpoint rule. The idea is to divide the time simulation into small segments. Then, one approximates the time evolution by the imaginary exponential of the Hamiltonian evaluated at the midpoint of each segment. As the number of segments increases, the evolution converges. This method may be identified with the 2nd-order Magnus method, i.e. $s = 1$. This is a special case however, in that the only error terms that contribute towards the overall bound are those due to truncating the Magnus Taylor series, using quadrature to approximate the exponential, and the subsequent Trotter product formula error.

\subsubsection{Suzuki method}

Instead of using a Trotter expansion of the Hamiltonian around the midpoint in each time segment, we can fit its time-ordered exponential.
This is the approach chosen by Refs.~\cite{wiebe2010higher,suzuki1990fractal,suzuki1991general,suzuki1993general}. Section IIB and Appendix A of~\cite{childs2021theory}, and Appendix V of~\cite{childs2019nearly} also discuss the error of applying similar strategies. We do not discuss its cost here as we already did in the main text.

\subsubsection{Continuous qDRIFT}

Both the CFQM operators and the Suzuki method rely on Suzuki-Trotter product formulas. A well-known variation called qDRIFT uses randomization to reduce the cost in some time-independent instances \cite{campbell2019random}. There exists similarly a time-dependent version of this method, called continuous qDRIFT \cite[Theorem 7']{berry2020time}. It inherits the same (dis)advantages and generalizes the prior Monte-Carlo-based method in Ref.~\cite{poulin2011quantum}. %Specifically, if we have $H(\tau) = \sum_{l=1}^L H_l(\tau)$, the evolution is probabilistically implemented with $\prod_{j=0}^{r-1}\mathcal{U}(t_j, t_{j+1})(\rho)$, where
%\begin{equation}
%    \mathcal{U}(t_j, t_{j+1})(\rho) := \sum_{l = 1}^L \int_{t_j}^{t_{j+1}} d\tau p_l(\tau) e^{-i\frac{H_l(\tau)}{p_l(\tau)}} \rho e^{i\frac{H_l(\tau)}{p_l(\tau)}},
%\end{equation}
%and $r \geq 4\left\lceil \frac{\|H_l(\tau)\|_{\infty,1,1}^2}{\epsilon}\right\rceil$. The norms in this last expression are taken over $H_l(\tau)$, the time, and index $l$, respectively~\cite[Theorem 7']{berry2020time}. 
Its main drawback is its scaling with the simulation time. If the norm of the Hamiltonian does not vary significantly over time, this indicates an implicit quadratic dependence, worse than our and previously discussed methods.

\subsubsection{Dyson series}

Last but not least, we discuss the Dyson series approach of Refs.~\cite{kieferova2019simulating,low2018hamiltonian}. In contrast to the above algorithms, it relies on an LCU: one expands the Dyson series in a similar fashion to the Taylor series approach and then uses oblivious amplitude amplification to ensure the success probability does not decrease~\cite{berry2015simulating}. The key distinction from the Taylor series approach is that one has to coherently order the times at which the Hamiltonian is evaluated.
%Let $H(t) =\sum_{l=1}^L \alpha_l(t) U_l$ be an LCU decomposition of $H(t)$. Let 
%\begin{equation}
%    U(0, t) = \mathcal{T}\left(\left[e^{-i\int_{0}^t d\tau H(\tau) }\right]\right)
%\end{equation}
%represent the time evolution operator, which can be rewritten as
%\begin{equation}
%\begin{split}
%    &U(0,t) = \sum_{k=0}^\infty \frac{(-i)^k}{k!}\int_0^{t}d\tau_1\cdots \int_0^{t}d\tau_k H(\tau'_1)\cdots H(\tau'_k)\\
%    &= \lim_{\substack{
%    K\rightarrow 0\\
%    M \rightarrow 0
%    }}\sum_{k=0}^K \frac{(-i)^k}{M^k k!}\sum_{m_1=0}^{M}\cdots \sum_{m_k=0}^{M} H\left(\frac{\tau m'_1}{M}\right)\cdots H\left(\frac{\tau m'_1}{M}\right).
%\end{split}
%\end{equation}
%The primed versions of $\tau_i$ and $m_i$ indicate the time-ordered variables. By appropriately truncating the series, we obtain a finite LCU that can be qubitized. 
The cost scaling of this method is given in Ref.~\cite[Theorem 10']{berry2020time},
\begin{equation}
O\left(L\|\alpha\|_{\infty,1}\frac{\log (L\|\alpha\|_{\infty,1}\epsilon^{-1})}{\log \log (L\|\alpha\|_{\infty,1}\epsilon^{-1})}\right)
\end{equation}
calls to the oracle that computes the coefficients, and $\tilde{O}(L^2\|\alpha\|_{\infty,1}g_c)$ additional gates, where $g_c$ is the cost of implementing each unitary in the linear combination. In the Heisenberg Hamiltonian from~\cref{eq:Heisenberg_Hamiltonian}, this translates into a complexity $\Omega(n)$ in the number of spins, similar to the complexity achieved by CFQMs and the Suzuki technique. The dependence on the error is more favorable here, however, though product formulas may achieve similar logarithmic error scaling using multi-product formulas, which we cover in \cref{app:Multi-product formulas}.   %The regimes where this LCU technique is less favorable than product-formula-based ones mirror the time-independent case. For example, if we take the unnormalized version of the Heisenberg Hamiltonian in \cref{eq:Heisenberg_Hamiltonian}, (e.g., without the factors of $(4n)^{-1}$), then the cost of the Dyson technique is $\Omega(n^2)$ \cite{haah2021quantum}, while the quasi-Magnus and Suzuki methods scale close to linearly in $n$ \cite{childs2019nearly}. In contrast, the cost scaling with the precision will be logarithmic, more favorable than the polynomial cost of product formulas. However, a similar logarithmic error scaling may be achieved using multi-product formulas, which we cover in \cref{app:Multi-product formulas}. 

\begin{figure*}[t]
    \centering
\includegraphics[width=0.35\textwidth]{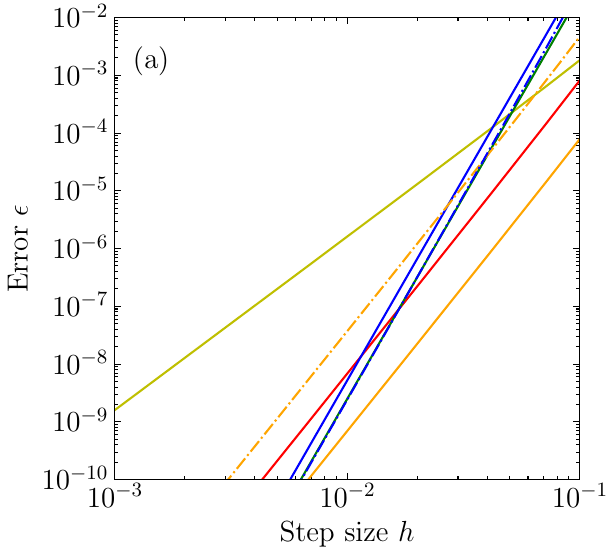}
\includegraphics[width=0.595\textwidth]{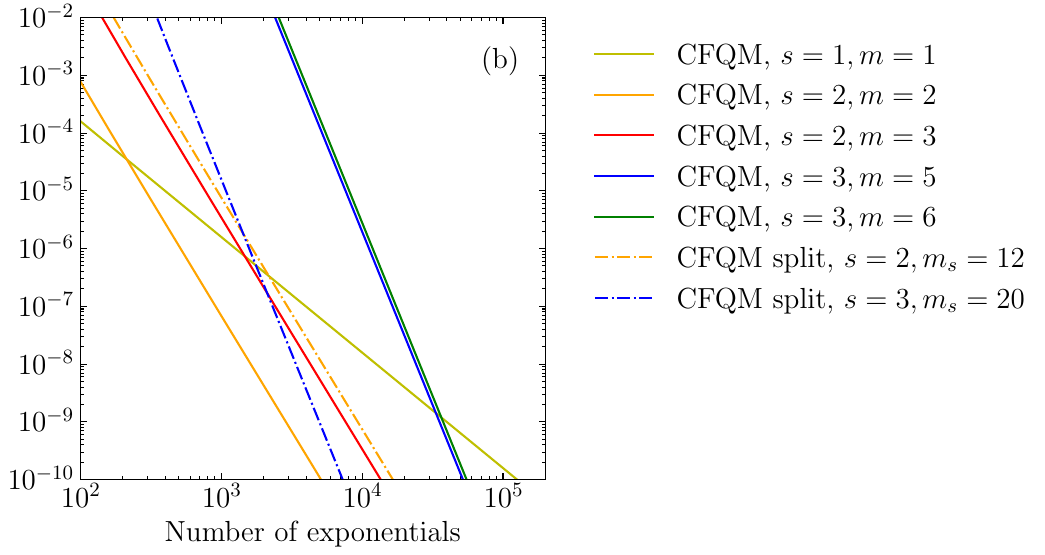}
    \caption{Relation of the error targeted $\epsilon$, and the step size $h$ and number of exponentials required to implement time evolution, as implied by our error bounds. We assume a Heisenberg Hamiltonian~\cref{eq:Heisenberg_Hamiltonian} of size $n = 128$. (a) Step error $\epsilon$ per step of size $h$. (b) Global error $\epsilon$ and number of exponentials incurred to time-evolve the Heisenberg Hamiltonian~\cref{eq:Heisenberg_Hamiltonian} for time $T = 1$.}
    \label{fig:step_error_cost}
\end{figure*}

\section{Proof of error bounds\label{app:proofs}}

In this section, we provide formal proofs for each of the error terms analyzed in this paper, corresponding to the first two error terms in~\cref{eq:general_error_bound_CFMagnus} (see also~\cref{fig:Magnus_derivation}). The first error term is further divided according to~\cref{eq:CFQM_definition_error}. Given the complicated expressions involved and the importance of rigorously capturing the constant factor of the complexity, we attempt to be as explicit as possible on the inequality steps we take.

\subsection{Taylor remainder of the Magnus operator\label{sec:Error_Taylor_Magnus}}
The first error we want to analyze is
\begin{equation}
    \sum_{p=2s+1}^\infty \left\|\exp(\Omega(h))_{p}\right\|.
\end{equation}
where $\exp(\Omega(h))_p$ denotes the order $O(h^p)$ expansion of $\exp(\Omega(h))$.
We recall
\begin{equation}
    A(t) = \sum_{i=0}^\infty a_i(t-t_{\frac{1}{2}})^{i}, \qquad a_i = \left.\frac{1}{i!}\frac{d^i A(t)}{dt^i}\right|_{t=t_{\frac{1}{2}}}.
\end{equation}
To bound the error from truncating the Taylor series to a given order, we use the Magnus operator expansion used in Refs.~\cite{mielnik1970combinatorial,moan2001convergence},
\begin{equation}
    \Omega(t_0, t_0+h) = \sum_{n=1}^\infty \int_{t_0}^{t_0+h} dt_n\ldots \int_{t_0}^{t_0+h} dt_1 L_n(t_n,\ldots, t_1) A(t_n)\ldots A(t_1).
\end{equation}
Here 
\begin{equation}
L_n(t_n,\ldots, t_1) = \frac{\Theta_n!(n-1-\Theta_n)!}{n!}(-1)^{n-1-\Theta_n},
\end{equation}
with $\Theta_n = \theta_{n-1,n-2} +\ldots +\theta_{2,1}$, and $\theta_{b,a}$ the step function that is valued $1$ when $t_b>t_a$ and $0$ otherwise. Consequently,
\begin{equation}
    |L_n(t_n,\ldots, t_1)|\leq \frac{(n-1)!}{n!} = \frac{1}{n}.
\end{equation}
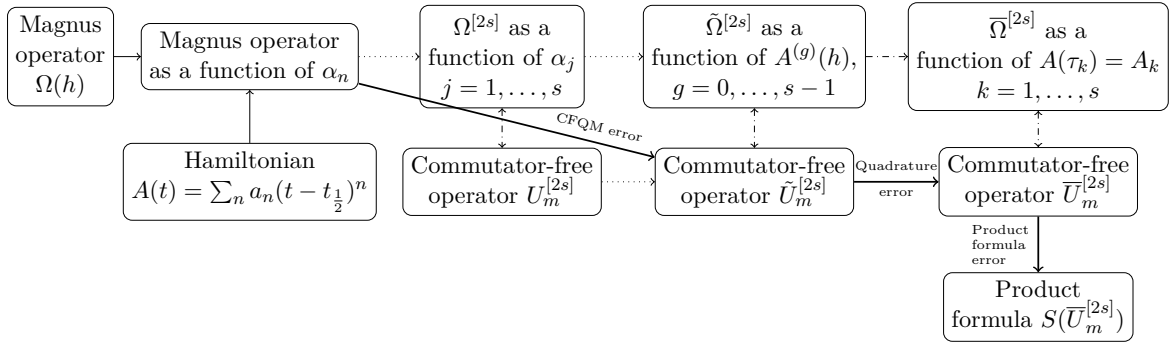
\begin{figure*}
    \centering
    \scalebox{0.83}{
    \begin{tikzpicture}[node distance=2cm, auto]

   % Node with more than one line and a name
  %\node[align=center, draw, rounded corners] (A) at (0,0) {$A(t)$};

  \node[align=center, draw, rounded corners] (A_Tay) at (3,0) {Hamiltonian \\$A(t) = \sum_n a_n (t-t_{\frac{1}{2}})^{n}$};

 %\draw[->] (A) -- (A_Tay);

  \node[align=center, draw, rounded corners] (Magnus) at (0,2) {Magnus \\ operator \\ $\Omega(h)$};

  \node[align=center, draw, rounded corners] (Magnus_Tay) at (3,2) {Magnus operator \\ as a function of $\alpha_n$};

  \draw[->] (A_Tay) -- (Magnus_Tay);
  \draw[->] (Magnus) -- (Magnus_Tay);

  \node[align=center, draw, rounded corners] (Omega2s) at (7,2) {$\Omega^{[2s]}$  as a\\ function of $\alpha_j$\\
  $j=1,\ldots,s$};

  \draw[->, dotted] (Magnus_Tay) -- (Omega2s);

  \node[align=center, draw, rounded corners] (Psi2s) at (7,0) {Commutator-free\\ operator $U_m^{[2s]}$};

  \draw[<->, dashdotted] (Omega2s) -- (Psi2s);

  \node[align=center, draw, rounded corners] (TildeOmega2s) at (11,2) {$\tilde{\Omega}^{[2s]}$ as a  \\ function of $A^{(g)}(h)$,\\ $g = 0,\ldots,s-1$};

  \draw[->, dotted] (Omega2s) -- (TildeOmega2s);

  \node[align=center, draw, rounded corners] (TildePsi2s) at (11,0) {Commutator-free\\ operator $\tilde{U}_m^{[2s]}$};

  \draw[->, dotted] (Psi2s) -- (TildePsi2s);

  \draw[<->, dashdotted] (TildeOmega2s) -- (TildePsi2s);

  \node[align=center, draw, rounded corners] (BarOmega2s) at (15.5,2) {$\overline{\Omega}^{[2s]}$ as a \\ function of $A(\tau_k) = A_k$\\ $k = 1,\ldots,s$};

  \draw[->, dashdotted] (TildeOmega2s) -- (BarOmega2s);

  \node[align=center, draw, rounded corners] (BarPsi2s) at (15.5,0) {Commutator-free\\ operator $\overline{U}_m^{[2s]}$};

  \draw[->, thick] (TildePsi2s) -- node[anchor=south, font=\tiny] {Quadrature} node[anchor=north, font=\tiny] {error} (BarPsi2s);

  %\draw [arrow] (dec1) -- node[anchor=east] {yes} (pro2a);

  \draw[<->, dashdotted] (BarOmega2s) -- (BarPsi2s);

  \node[align=center, draw, rounded corners] (Trotter) at (15.5,-2) {Product \\ formula $S(\overline{U}_m^{[2s]})$};

    \draw[->, thick] (Magnus_Tay) -- node[pos=0.8, sloped, above, font=\tiny] {CFQM error} (TildePsi2s);

    \draw[->, thick] (BarPsi2s) -- node[anchor=east, font=\tiny, align=left] {Product \\
    formula \\
    error} (Trotter);

    \end{tikzpicture}
    }
\caption{\label{fig:Magnus_derivation}Flowchart indicating how the commutator-free quasi-Magnus (CFQM) operators can be defined and implemented; and the errors corresponding to the approximations made along the way. The dashdotted lines indicate equality up to $O(h^{2s})$; whereas the dotted ones indicate a formal identification between two Lie algebras, graded (expanded by $\{\alpha_j\}$) and ungraded (expanded by $\{A_k\}$ or $\{A^{(g)}(h)\}$). The three solid thick lines represent the three error sources we evaluate. Alternative derivations of CFQMs are described in Refs.~\cite{alvermann2011high} and~\cite{blanes2017high}, based on Legendre and Lagrange polynomials respectively.}
\end{figure*}
Then, we expand
\begin{align}
    &\Omega(t_0, t_0+h) = \sum_{n=1}^\infty \int_{t_0}^{t_0+h} dt_n\ldots \int_{t_0}^{t_0+h} dt_1 L_n(t_n,\ldots, t_1) A(t_n)\ldots A(t_1)\\
    &= \sum_{n=1}^\infty \int_{t_0}^{t_0+h} dt_n\ldots \int_{t_0}^{t_0+h} dt_1 L_n(t_n,\ldots, t_1)\left(\sum_{i=0}^\infty a_i(t_n-t_{\frac{1}{2}})^{i}\right)\ldots \left(\sum_{i=0}^\infty a_i(t_1-t_{\frac{1}{2}})^{i}\right)\\
    &= \sum_{n=1}^\infty\sum_{p=0}^\infty \sum_{p_1+\ldots+p_n = p}(a_{p_1}\cdot\ldots\cdot a_{p_n}) \int_{t_0}^{t_0+h} dt_n\ldots \int_{t_0}^{t_0+h} dt_1 L_n(t_n,\ldots, t_1)(t_1-t_{\frac{1}{2}})^{p_1}\cdot\ldots\cdot(t_n-t_{\frac{1}{2}})^{p_n}.
\end{align}

Using the latter expression, we have the following estimate of all orders $p$
\begin{align}
    \|\Omega(t_0, t_0+h)\| &\leq \sum_{n=1}^\infty\sum_{k=0}^\infty \sum_{p_1+\ldots+p_n = p}\bigg(\|a_{p_1}\cdot\ldots\cdot a_{p_n}\|\cdot \\
    & \left.\int_{t_0}^{t_0+h} dt_n\ldots \int_{t_0}^{t_0+h} dt_1 |L_n(t_n,\ldots, t_1)|\cdot|t_1-t_{\frac{1}{2}}|^{p_1}\cdot\ldots\cdot|t_n-t_{\frac{1}{2}}|^{p_n}\right)\\
    \leq \sum_{n=1}^\infty\frac{1}{n}&\sum_{p=0}^\infty \sum_{p_1+\ldots+p_n = p}\|a_{p_1}\cdot\ldots\cdot a_{p_n}\| \left(\int_{t_0}^{t_0+h} dt_n|t_n-t_{\frac{1}{2}}|^{p_n}\right)\ldots \left( \int_{t_0}^{t_0+h} dt_1 |t_1-t_{\frac{1}{2}}|^{p_1}\right)\\
    &= \sum_{n=1}^\infty\frac{1}{n}\sum_{p=0}^\infty \sum_{p_1+\ldots+p_n = p}\|a_{p_1}\cdot\ldots\cdot a_{p_n}\| \left(\frac{2\left(\frac{h}{2}\right)^{p_n+1}}{p_n+1}\right)\ldots \left(\frac{2\left(\frac{h}{2}\right)^{p_1+1}}{p_1+1}\right)\\
    &= \sum_{n=1}^\infty\sum_{p=0}^\infty \frac{2^n\left(\frac{h}{2}\right)^{p+n}}{n} \sum_{p_1+\ldots+p_n = p} \left\|\frac{a_{p_n}}{p_n+1}\right\|\ldots \left\|\frac{a_{p_1}}{p_1+1}\right\|.
\end{align}
Consequently,
\begin{align}\label{eq:bound_Magnus_order}
    \|\Omega(h)\|&\leq \sum_{n=1}^\infty\sum_{p=0}^\infty \frac{h^{p+n}}{n2^{p+n}} \sum_{p_1+\ldots+p_n = p} \left\|\frac{2a_{p_n}}{p_n+1}\right\|\ldots \left\|\frac{2a_{p_1}}{p_1+1}\right\|,
\end{align}
where $p_1,\ldots,p_n\in\mathbb{N}$.
We rename them $j_i = p_i+1$ and also $k = p+n$ and the upper bound of $k$ becomes $2s+n$ in the previous equation.
Then,
\begin{align}
    \|\exp(\Omega(h))\|&\leq \sum_{z=0}^\infty\frac{1}{z!}\left(\sum_{n=1}^\infty\sum_{k=n}^\infty \frac{h^{k}}{n2^{k}} \sum_{j_1+\ldots+j_n = k} \left\|\frac{2a_{j_n-1}}{j_n}\right\|\ldots \left\|\frac{2a_{j_1-1}}{j_1}\right\|\right)^z\\
    &= \sum_{z=0}^\infty\frac{1}{z!} \prod_{l=1}^z\sum_{n_l=1}^\infty\sum_{k_l=n_l}^\infty\frac{h^{k_l}}{n_l2^{k_l}} \sum_{{j_1}_l+\ldots+j_{n_l} = k_l} \left\|\frac{2a_{j_{n_l}-1}}{j_{n_l}}\right\|\ldots \left\|\frac{2a_{{j_1}_l-1}}{{j_1}_l}\right\|\\
    &= \sum_{z=0}^\infty\frac{1}{z!} \sum_{n_1, \ldots,n_z=1}^\infty\sum_{\substack{
    k_1,\ldots,k_z\\
    =n_1,\ldots,n_l
    }}^\infty 
    %\sum_{
    %\substack{
    %{j_1}_1+\ldots+j_{n_1} = k_1\\
    %\ldots\\
    %{j_1}_z+\ldots+j_{n_z} = k_z
    %}}
    \prod_{l=1}^z \frac{h^{k_l}}{n_l2^{k_l}}
    \sum_{{j_1}_l+\ldots+j_{n_l} = k_l} 
    \left\|\frac{2a_{j_{n_l}-1}}{j_{n_l}}\right\|\ldots \left\|\frac{2a_{{j_1}_l-1}}{{j_1}_l}\right\|\\
    &=\sum_{p=0}^\infty\left(\frac{h}{2}\right)^p\sum_{z=0}^\infty\frac{1}{z!}\sum_{n_1, \ldots,n_z=1}^\infty\sum_{\substack{\sum_{i=1}^z k_i=p\\k_i \geq n_i}}
    \prod_{l=1}^z \frac{1}{n_l} \sum_{{j_1}_l+\ldots+j_{n_l} = k_l} \left\|\frac{2a_{j_{n_l}-1}}{j_{n_l}}\right\|\ldots \left\|\frac{2a_{{j_1}_l-1}}{{j_1}_l}\right\|.
\end{align}
Now, we bound $\|a_{j}\|\leq c$, and denote $\mathcal{C}(p)$ the compositions of $p$, as well as $\mathcal{C}^z(p)$ the compositions of size $z$. If we write the last summatory like a sum over $\bm{j}_l\in\mathcal{C}^{n_l}(k_l)$, as $j_i = p_i+1>0$, the condition $k_i\geq n_i$ becomes redundant. Therefore we can further sum over all the $n_l$ to get $\bm{j}\in\mathcal{C}(k_l)$.
Consequently, we write
\begin{align}\label{eq:exp_Omega_intermediate_bound}
\|\exp(\Omega(h))\|&= \sum_{p=0}^\infty\left(\frac{h}{2}\right)^p\sum_{z=0}^\infty \frac{1}{z!}\sum_{\bm{k}\in\mathcal{C}^z(p)}
    \prod_{l=1}^z
    \sum_{\substack{
    \bm{j}_l \in \mathcal{C}(k_l)
    }}\frac{1}{\dim \bm{j}_l}\prod_{\ell_l=1}^{\dim \bm{j}_l}  
    \frac{2\|a_{j_{n_l}-1}\|}{j_{\ell_l}}\\
&\leq\sum_{p=0}^\infty\left(\frac{h}{2}\right)^p\sum_{z=0}^\infty \frac{1}{z!}\sum_{\bm{k}\in\mathcal{C}^z(p)}  \prod_{l=1}^z
    \sum_{\substack{
    \bm{j}_l \in \mathcal{C}(k_l)
    }}\frac{(2c)^{\dim \bm{j}_l}}{\dim \bm{j}_l}
    \prod_{\ell_l=1}^{\dim \bm{j}_l}  
    \frac{1}{j_{\ell_l}}\\
    &= \sum_{p=0}^\infty\left(\frac{h}{2}\right)^p \sum_{\bm{k}\in\mathcal{C}(p)}\frac{1}{(\dim \bm{k})!} \prod_{l=1}^{\dim \bm{k}} \sum_{\bm{j}_l\in\mathcal{C}(k_l)} \frac{(2c)^{\dim \bm{j}_l}}{\dim \bm{j}_l}\prod_{\ell_l=1}^{\dim \bm{j}_l}  
    \frac{1}{j_{\ell_l}}.
\end{align}
Overall, this means that
\begin{align}\label{eq:exp_Omega_error}
    \sum_{p=2s+1}^\infty \|\exp(\Omega)_{p}\| \leq \sum_{p=2s+1}^\infty\left(\frac{h}{2}\right)^{p} \sum_{\bm{k}\in\mathcal{C}(p)}\frac{1}{(\dim \bm{k})!} \prod_{l=1}^{\dim \bm{k}} \sum_{\bm{j}_l\in\mathcal{C}(k_l)} \frac{(2c)^{\dim \bm{j}_l}}{\dim \bm{j}_l}\prod_{\ell_l=1}^{\dim \bm{j}_l}  
    \frac{1}{j_{\ell_l}}.
\end{align}

\subsection{Taylor remainder of the quasi-Magnus operator \label{sec:Error_Taylor_quasiMagnus}}
The goal of this section is to bound
\begin{equation}
    \sum_{p = 2s+1}^{\infty}\|\tilde{U}_{m,p}^{[2s]}\|.
\end{equation}
We know that $\tilde{U}_{m}^{[2s]}$ will be correct up to order $O(h^{2s+1})$. Still, we need to figure out an error bound to this factor. To do so, we expand
\begin{align}
    \tilde{U}_{m}^{[2s]} = \prod_{i=1}^m \exp\left(\sum_{g=0}^{s-1} y_{i,g} A^{(g)}(h)\right) = \prod_{i=1}^m \left(\sum_{k=0}^\infty\frac{(\sum_{g=0}^{s-1} y_{i,g} A^{(g)}(h))^k}{k!}\right).
\end{align}
Recall that $A^{(g)}(h) = \sum_{j=1}^\infty\frac{1-(-1)^{g+j}}{(g+j)2^{g+j}}\alpha_j$ and $\alpha_j = \frac{1}{(j-1)!}\left.\frac{d^{j-1} A}{dt^{j-1}}\right|_{t = t_{\frac{1}{2}}}h^j$ contains a factor of $h^j$.
Now we convert the product of exponentials into a plain sum, starting with
\begin{align}
    \sum_{g=0}^{s-1} y_{i,g} A^{(g)}(h) &= \sum_{g=0}^{s-1} y_{i,g} \sum_{j=1}^\infty\frac{1-(-1)^{g+j}}{(g+j)2^{g+j}}\alpha_j \\
    &=\sum_{j=1}^\infty \sum_{g=0}^{s-1} y_{i,g} \frac{1-(-1)^{g+j}}{(g+j)2^{g+j}}\alpha_j .
\end{align}
Let us now call 
\begin{equation}
    \overline{x}_{i,j} := \sum_{g=0}^{s-1} y_{i,g} \frac{1-(-1)^{g+j}}{(g+j)2^{g+j}}
\end{equation}
which is an extension of $x_{i,j}$ for values of $j>s$. Then, we can write
\begin{align}
    \sum_{g=0}^{s-1} y_{i,g} A^{(g)}(h) &= \sum_{j=1}^\infty \overline{x}_{i,j} \alpha_j 
\end{align}
and
\begin{align}
    \left(\sum_{g=0}^{s-1} y_{i,g} A^{(g)}(h)\right)^k &= \left(\sum_{j=1}^\infty \overline{x}_{i,j} \alpha_j\right)^k  = \sum_{j_1,\ldots, j_k\in(\mathbb{Z}^+)^k} \prod_{l\in \{1,\ldots,k\}} \overline{x}_{i,j_l} \alpha_{j_l}.
\end{align}
Substituting back,
\begin{align}
    \tilde{U}_{m}^{[2s]} &= \prod_{i=1}^m \exp\left(\sum_{g=0}^{s-1} y_{i,g} A^{(g)}(h)\right) = \prod_{i=1}^m \left(\sum_{k=0}^\infty\frac{(\sum_{g=0}^{s-1} y_{i,g} A^{(g)}(h))^k}{k!}\right)\\
    &= \prod_{i=1}^m \left(\sum_{k=0}^\infty\frac{1}{k!}\sum_{j_1,\ldots, j_k\in(\mathbb{Z}^+)^k} \prod_{l\in \{1,\ldots,k\}}\overline{x}_{i,j_l} \alpha_{j_l}\right)\\
    &= \sum_{k_1,\ldots,k_m=0}^\infty \frac{1}{k_1!\ldots k_m!} \sum_{\substack{
    j_{1,1},\ldots, j_{1,k_1}\in(\mathbb{Z}^+)^{k_1}\\
    \cdots \\
    j_{m,1},\ldots, j_{m,k_m}\in(\mathbb{Z}^+)^{k_m}}
    }
    \prod_{\substack{
    i\in \{1,\ldots,m\}
    }}\prod_{l_i\in \{1,\ldots,k_i\}}
    \overline{x}_{i,j_{l_i}} \alpha_{j_{l_i}}.
\end{align}

Let us assume that $\|\overline{x}_{i,j_l}\alpha_{j_l}\| = \|\overline{x}_{i,j_l}a_{j_l-1}\|h^{j_l} \leq \bar{c} h^{j_l}$. Thus,
\begin{align}
    \|\tilde{U}_{m}^{[2s]}\| &\leq\sum_{k_1,\ldots,k_m=0}^\infty \frac{1}{k_1!\ldots k_m!} \sum_{\substack{
    j_{1,1},\ldots, j_{1,k_1}\in(\mathbb{Z}^+)^{k_1}\\
    \ldots \\
    j_{m,1},\ldots, j_{m,k_m}\in(\mathbb{Z}^+)^{k_m}}
    }
    \prod_{\substack{
    i\in \{1,\ldots,m\}
    }}\prod_{l_i\in \{1,\ldots,k_i\}}
   \| \overline{x}_{i,j_{l_i}} \alpha_{j_{l_i}}\| \\
    &\leq \sum_{k_1,\ldots,k_m=0}^\infty \frac{1}{k_1!\ldots k_m!} \sum_{\substack{
    j_{1,1},\ldots, j_{1,k_1}\in(\mathbb{Z}^+)^{k_1}\\
    \ldots \\
    j_{m,1},\ldots, j_{m,k_m}\in(\mathbb{Z}^+)^{k_m}}
    }
    \prod_{\substack{
    i\in \{1,\ldots,m\}
    }}\bar{c}^{k_i} h^{\sum_{l_i\in\{1,\ldots,k_i\}} j_{l_i}}\\
    &= \sum_{k_1,\ldots,k_m=0}^\infty \sum_{\substack{
    j_{1,1},\ldots, j_{1,k_1}\in(\mathbb{Z}^+)^{k_1}\\
    \ldots \\
    j_{m,1},\ldots, j_{m,k_m}\in(\mathbb{Z}^+)^{k_m}}
    } 
    \frac{\bar{c}^{k_1}\cdots  \bar{c}^{k_m} }{k_1!\cdots k_m!} h^{\sum_{i=1}^m\sum_{l_i\in\{1,\ldots,k_i\}} j_{l_i}}.
\end{align}
We now need to group the terms by the exponent of $h$. Let $\bm{w} = [j]$ be an element of $\mathcal{C}(p)$ of size $\dim\bm{w}$. Then, there are $m^{\dim\bm{w}}$ ways of allocating the terms of $\bm{w}$ between $k_1,\ldots,k_m$. Consequently, we can bound the term of $\tilde{U}_{m}^{[2s]}$ of order $O(h^p)$, $\tilde{U}_{m,p}^{[2s]}$, by
\begin{equation}
    \|\tilde{U}_{m,p}^{[2s]}\| \leq h^p \sum_{
        \substack{
            \bm{w}\in\mathcal{C}(p)
            }
    }
    \sum_{k_i : \sum_{i=1}^m k_i = \dim\bm{w}}
    \frac{\bar{c}^{k_1}\cdots \bar{c}^{k_m}}{k_1!\cdots k_m!} 
    = h^p \sum_{
        \substack{
            \bm{w}\in\mathcal{C}(p)
            }
    }
    \bar{c}^{\dim\bm{w}}
    \sum_{k_i : \sum_{i=1}^m k_i = \dim\bm{w}}
    \frac{1}{k_1!\cdots k_m!}.
\end{equation}
The expression $k_i : \sum_{i=1}^m k_i = \dim\bm{w}$ can be understood as the \textit{weak} integer compositions of $\dim\bm{w}$. Overall, we will bound the quasi-Magnus Taylor error as
\begin{equation}\label{eq:Psi_m_Taylor_error} \sum_{p=2s+1}^\infty\|\tilde{U}_{m,p}^{[2s]}\| \leq \sum_{p=2s+1}^\infty h^{p} \sum_{
        \substack{
            \bm{w}\in\mathcal{C}(p)
            }
    }
    \sum_{k_i : \sum_{i=1}^m k_i = \dim\bm{w}}
    \frac{\bar{c}^{k_1}\cdots \bar{c}^{k_m}}{k_1!\cdots k_m!}.
\end{equation}

\subsection{Quadrature error\label{sec:Quadrature_error}}

We want to compute the error 
\begin{equation}
    \left\|\tilde{U}_m^{[2s]} -\overline{U}_m^{[2s]}\right\| = \left\| \prod_{i=1}^m \exp\left(\sum_{g=0}^{s-1} y_{i,g} A^{(g)}(h)\right) - \prod_{i=1}^m \exp\left(\sum_{k=1}^s z_{i,k} A_k h\right)
 \right\|.
\end{equation}
For the sake of simplicity, let us call the exponents $X_1,\ldots,X_m$ and $Y_1,\ldots,Y_m$. Then
\begin{align}
    e^{X_1}\ldots e^{X_m} &= (e^{Y_1}+e^{X_1}-e^{Y_1})e^{X_2}\ldots e^{X_m} = e^{Y_1}e^{X_2}\ldots e^{X_m} + (e^{X_1}-e^{Y_1})e^{X_2}\ldots e^{X_m}\\
    &= e^{Y_1}e^{Y_2}\ldots e^{X_m} + (e^{X_1}-e^{Y_1})e^{X_2}\ldots e^{X_m} + e^{Y_1}(e^{X_2}-e^{Y_2})e^{X_2}\ldots e^{X_m}\\
    &\ldots\\
    &= e^{Y_1}\ldots e^{Y_m} + \sum_{j=1}^m e^{Y_1}\ldots e^{Y_{j-1}} (e^{X_j}-e^{Y_j}) e^{X_{j+1}}\ldots e^{X_{m}}.
\end{align}
Since all operators $e^{X_i}$ and $e^{Y_i}$ are unitary, 
\begin{align}
   \|e^{X_1}\cdots e^{X_m} - e^{Y_1}\cdots e^{Y_m}\|&\leq \sum_{j=1}^m \|e^{Y_1}\|\cdots \|e^{Y_{j-1}}\| \cdot\|e^{X_j}-e^{Y_j} \| \cdot\|e^{X_{j+1}}\|\cdots \|e^{X_{m}}\| \\
   &=\sum_{j=1}^m \|e^{X_j}-e^{Y_j} \|.
\end{align}
Now we apply the following corollary:
\begin{cor}[Corollary A.5 in \cite{childs2021theory}]
Given $\mathcal{H}$ and $\mathcal{G}$ two antihermitian continuous operator-valued functions defined on $\mathbb{R}$,
\begin{equation}
\left\|\mathcal{T}\exp\left(\int_{t_1}^{t_2}d\tau \mathcal{H}(\tau)\right)-\mathcal{T}\exp\left(\int_{t_1}^{t_2}d\tau \mathcal{G}(\tau)\right)\right\|\leq \left|\int_{t_1}^{t_2}d\tau \|\mathcal{H}(\tau)-\mathcal{G}(\tau)\| \right|.
\end{equation}
\end{cor}
Consequently,
\begin{align}
    \|e^{X_1}\cdots e^{X_m} - e^{Y_1}\cdots e^{Y_m}\| \leq \sum_{j=1}^m \|X_j-Y_j\|.
\end{align}

Now we focus on a single exponential, $X_i = \sum_{g=0}^{s-1}y_{i,g} A^{(g)}(h)$. The error from approximating the univariate integral $A^{(g)}$ with quadrature
\begin{equation}
    \int_a^b f(x) dx = \sum_{k=1}^{n}w_k f(c_k)+ R_n
\end{equation}
can be bounded using the expression~\cite[Pag. 146, Chapter 5]{kahaner1989numerical}
\begin{equation}
    R_n = \frac{(b-a)^{2n+1}(n!)^4}{(2n+1)((2n)!)^3}f^{(2n)}(\xi),\qquad a<\xi<b.
\end{equation}
As a consequence, the error from approximating~\eqref{eq:A^i_integrals} with $n$-point Gauss-Legendre quadrature is
\begin{align}\label{eq:quadrature_residual}
\|R_n(A^{(g)})(h)\| &= \frac{h^{2n+1}(n!)^4}{(2n+1)((2n)!)^3}\left\|\left.\frac{d^{2n}}{dt^{2n}}\frac{1}{h^g} \sum_{j = 0}^\infty a_j  t^{g+j} \right|_{t=\xi}\right\|\\
&= \frac{h^{2n+1}(n!)^4}{(2n+1)((2n)!)^3}\frac{1}{h^g} \left\|\sum_{j = 0}^\infty a_j\left.\frac{d^{2n}}{dt^{2n}}  t^{g+j} \right|_{t=\xi}\right\|\\
&= \frac{h^{2n+1}(n!)^4}{(2n+1)((2n)!)^3}\frac{1}{h^g}\left\|\sum_{j = 2n-g}^\infty a_j\left.\frac{(g+j)!}{(g+j-2n)!}t^{g+j-2n}\right|_{t=\xi}\right\|\\
&= \frac{h^{2n+1}(n!)^4}{(2n+1)((2n)!)^3}\frac{1}{h^g} \left\|\sum_{j = 0}^\infty a_{j+2n-g}\left.\frac{(j+2n)!}{j!}t^{j}\right|_{t=\xi}\right\|\\
&\leq \frac{h^{2n+1}(n!)^4}{(2n+1)((2n)!)^3}\frac{1}{h^g} \sum_{j = 0}^\infty \frac{(j+2n)!}{j!}\frac{h^j}{2^j}\|a_{j+2n-g}\|.
\end{align}
Since the integral limits $a$ and $b$ are $\pm \frac{h}{2}$ in the second equality in~\eqref{eq:A^i_integrals}, $|\xi|<\frac{h}{2}$. Then, we can use this expression to bound the error
\begin{equation}
    \left\| \tilde{U}_{m}^{[2s]} -\overline{U}_m^{[2s]} \right\| \leq \sum_{i=1}^m\sum_{g=0}^{s-1} \left\|y_{i,g} R_{s}(A^{(g)})\right\|.
\end{equation}

\section{Split-operator CFQMs\label{app:split_operator}}
There is an alternative pathway to obtain a CFQM. If the Hamiltonian has the form $-i H(t) = T(t)+V(t)$, where $[T(t),T(t')] = [V(t),V(t')]=0, \forall t,t'$,  we can expand the Magnus operator as a product formula (Refs.~\cite{blanes2006splitting} and~\cite[Sec. 5.8.2]{blanes2009magnus})
\begin{align}\label{eq:SO_definition_generators}
    W^{[2s]}_{m_s} = \prod_{i=1}^{m_s} e^{\tilde{T}_i} e^{\tilde{V}_i},\qquad
    \tilde{T}_i = \sum_{j=1}^s \bar{t}_{ij}\theta_j,\qquad \tilde{V}_i = \sum_{j=1}^s \bar{v}_{ij}\nu_j,
\end{align}
where $\theta_j$ and $\nu_j$ are Lie algebra generators akin to $\alpha_j$. Further, the parameters $\bar{t}_{ij}$ and $\bar{v}_{ij}$ fulfill the same time anti-symmetry constraint ~\eqref{eq:time_symmetry_parameter_constraint} as $x_{ij}$.
Similar to before, the process in finding the coefficients starts by expanding the operators into the Taylor series
\begin{align}\label{eq:Taylor_split_V}
    V(t) = \sum_{j= 0}^\infty v_j (t-t_{\frac{1}{2}})^j, \qquad v_j = \left.\frac{1}{j!}\frac{d^j V(t)}{dt^j}\right|_{t=t_{\frac{1}{2}}}, \\
\label{eq:Taylor_split_T}
    T(t) = \sum_{j= 0}^\infty w_j (t-t_{\frac{1}{2}})^j, \qquad w_j = \left.\frac{1}{j!}\frac{d^j T(t)}{dt^j}\right|_{t=t_{\frac{1}{2}}},
\end{align}
and substitute the expansions in the Magnus series recursion, obtaining an expression similar to~\cref{eq:Omega2_1generators,eq:Omega4_2generators,eq:Omega6_3generators}. Then, we take the two expressions we want to make equal, i.e. $\exp(\Omega^{[2s]})$ and $W^{[2s]}_{m_s}$, expand in Taylor series of $h$ to the appropriate order, and choose the parameters $\bar{t}_{ij}$ and $\bar{v}_{ij}$ to fit the coefficients in the Taylor series of the Magnus operator. Finally, we express $v_i,w_i$ in terms of univariate integrals, which we would write as $\tilde{W}^{[2s]}_{m_s}$ and, using an appropriate order quadrature to reproduce such integrals, arrive at
\begin{equation}
    \overline{T}_i = \sum_{k=1}^s \rho_{ik} T_k h,\qquad \overline{V}_i = \sum_{k=1}^s
    \sigma_{ik} V_k h,
\end{equation}
with
\begin{equation}\label{eq:split_variable_change}
    R^{(m,k)} = A^{(m,s)}R^{(s)}Q^{(s,k)}, \qquad  S^{(m,k)} = B^{(m,s)}R^{(s)}Q^{(s,k)}
\end{equation}
for $(R^{(m,k)})_{ij}=\rho_{ij}$ and $(S^{(m,k)})_{ij}=\sigma_{ij}$. Matrices $A^{(m,s)}$ and $B^{(m,s)}$ contain the parameters we need to find, $\overline{t}_{ij}$ and $\overline{v}_{ij}$ respectively. $R^{(s)}$ is the inverse of $T^{(s)}$ defined in~\eqref{eq:A^i_integrals}, while $Q^{(s,k)}$ is the quadrature matrix
\begin{equation}\label{eq:Gauss_Legendre_quadrature}
    Q^{(s,k)} = \begin{pmatrix}
        w_1 & \ldots & w_k\\
        \vdots & \ddots & \vdots\\
        w_1c_1^{s-1} & \ldots & w_k c_k^{s-1}
    \end{pmatrix},
\end{equation}
with $w_k$ and $c_k\in(-1,1)$ the weights and nodes of the quadrature.
Consequently, we end up implementing
\begin{equation}\label{eq:SO_definition_quadrature}
    \overline{W}^{[2s]}_{m_s} =\prod_{i=1}^{m_s} \left( \exp\left[{\sum_{k=1}^s \rho_{ik} T_k} h\right] \exp\left[\sum_{k=1}^s \sigma_{ik} V_k h\right]\right).
\end{equation}
Since operators $T_k$ and $V_k$ commute with themselves at different times, we can fast-forward the resulting exponentials by implementing the individual evolutions $\exp(\sum_k\rho_{ik} T_k h)$ and $\exp(\sum_k\sigma_{ik} V_k h)$ without the need for Trotter product formula decompositions.

Let us now analyze how to bound the error of this variation. Clearly, we will no longer need the product formula error as the Hamiltonian is already split. This leaves us with the other two terms. The quadrature error uses the same formula, the only adaptation is that the norm $\|a_j\|$ of Hamiltonian $A(t)$ and its derivatives will be substituted by those of its components $T$ and $V$, $\|v_j\|$ and $\|w_j\|$ according to~\eqref{eq:Taylor_split_V} and~\eqref{eq:Taylor_split_T}. Similarly $y_j^{(i)}$ will have to be substituted by the corresponding coefficients of the univariate integrals obtained from $A^{(m,s)}R^{(s)}$ and $B^{(m,s)}R^{(s)}$, see~\eqref{eq:split_variable_change}. Finally, for both Taylor expansion errors discussed in~\cref{sec:Error_Taylor_Magnus} and~\cref{sec:Error_Taylor_quasiMagnus} there are two key changes. First, the number of exponentials $m$ grows significantly, and we will denote it as $m_s$ in the numerical results in the next section to highlight that its larger number is due to using the split operator approach. And second, the norm $c$ and $\bar{c}$ now refer to the upper bounds of the norm of individual components.

The two split-operator commutator-free quasi-Magnus operators whose cost we analyze in this work are the order 4 and order 6 methods presented under the name GS$_{6}$-$4$ and GS$_{10}$-$6$ introduced in tables 2 and 3 of Ref.~\cite{blanes2006splitting}. GS stands for general splitting, which applies to the Hamiltonian discussed here, e.g. $H(t) = T(t) + V(t)$ with $T(t)$ and $V(t)$ commuting with themselves at different times. This condition applies to certain spin models as well as electronic structure Hamiltonians decomposing to the kinetic and Coulomb potentials.

We have carried out the analysis for a Hamiltonian with two fast-forwardable terms. However, given the equivalent expression to~\eqref{eq:SO_definition_quadrature} with more fast-forwardable terms, we could similarly apply our error bounds as these only depend on the number of exponentials and bounds to the norm of the Hamiltonian terms and their derivatives.

\section{Multi-product formulas\label{app:Multi-product formulas}}
One option to further optimize the cost of the Magnus simulation of time-dependent Hamiltonian is to replace the resulting product formula with a multi-product formula~\cite{childs2012LCUs}. Such multi-product formulas are a linear combination of powers of product formulas $U_{2s}\left(H;h\right)$, of the form
\begin{equation}
    U_{2s,\vec{k}}(H; h) = \sum_{j=1}^M a_j U_{2s}^{k_j}\left(H;\frac{h}{k_j}\right) = e^{-ihH} + O(h^{2s'+1}),
\end{equation}
where the first argument indicates the Hamiltonian, and the second the segment lengths. For us, $U_{2s}\left(H;\frac{h}{k_j}\right)$ would be the CFQM $S_{2s}\left(\overline{U}_m^{[2s]}\left( \frac{h}{k_j}\right)\right)$ and $M$ can be chosen to be equal to $s'\geq s$~\cite{low2019well}.
It is also possible to select the parameters $k_j$ and amplitudes $a_j$ such that $\|\vec{k}\|_{1} = O(s'^2\log s')$, while the success probability, $O(\|\vec{a}\|_1^{-1})$ with $\|\vec{a}\|_1 = O(\log s')$, can be easily amplified to $O(1)$ via oblivious amplitude amplification~\cite{berry2014exponential,berry2015simulating,low2019well}. Overall, this means that with a multiplicative overhead of $O(s'^2\text{poly}\log s')$ over the cost of standard product formulas, multi-product variations exponentially reduce the error from $O(h^{2s+1})$ to $O(h^{2s'+1})$. In consequence, multi-product formulas achieve an exponential cost reduction (in terms of target accuracy) compared to the alternative of increasing the order of the product formula, where each 2-order increase multiplies the cost by 5,~\cite{childs2021theory}.

The key requirement for the multi-product construction procedure in Ref.~\cite{low2019well} to work, is that we can Taylor expand the exact evolution as~\cite{blanes1999extrapolation,blanes2005raising,chin2010multi}
\begin{equation}
    \left[S_{2s}\left(\overline{U}_m^{[2s]}\left( \frac{h}{k_j}\right)\right)\right]^{k_j}= e^{\Omega(h)} + \frac{h^{2s+1}}{k_j^{2s}}\tilde{E}_{2s+1}(h) + \frac{h^{2s+3}}{k_j^{2s+2}}\tilde{E}_{2s+3}(h) +\ldots,
\end{equation}
with $\tilde{E}_{2s+i}(h)$ representing arbitrary error operators. A sufficient condition for this to happen is that
\begin{equation}S_{2s}\left(\overline{U}_m^{[2s]}\left( h\right)\right) = e^{\Omega(h) + h^{2s+1}E_{2s+1} + h^{2s+3}E_{2s+3} +\ldots},
\end{equation}
with $E_{2s+i}$ again representing error operators, constant in $h$.
This requirement is equivalent to stating that only odd terms in $h$ will be missing from the exact expression of $\Omega(h)$. It is clear that this same structure is followed in the truncation of the Magnus operator in~\cref{eq:Omega2_1generators,eq:Omega4_2generators,eq:Omega6_3generators}: only odd terms appear in the expression, due to the time anti-symmetry that we enforced, $\Omega(t_0, t_0+h) = - \Omega(t_0+h, t_0)$~\cite{blanes2000improved}. In this case, the error operators will be given by the missing commutators of the derivatives of the Hamiltonian evaluated at $t_{\frac{1}{2}}$, the $a_j$ operators.
Thus, we will argue that an operator fulfills the requirements to implement multi-product formulas whenever it exhibits the time anti-symmetric behavior.

Such time anti-symmetry is also enforced in the commutator-free quasi-Magnus operators thanks to the condition that $x_{m+1-i,j} = (-1)^{j+1}x_{i,j}$, see~\cref{eq:time_symmetry_parameter_constraint}. This condition will similarly carry over to the expression of $\tilde{U}_m^{[2s]}$. We can see this is the case due to two key facts. First, 
\begin{align}
    \equalto{\sum_j x_{i,j}\alpha_j}{} &= \sum_{jg}y_{i,g} T_{g,j}^{(s)}\alpha_j = \sum_g y_{i,g} A^{(g)}(h)\\
    \sum_j (-1)^{j+1} x_{m+1-i,j}\alpha_j &= \sum_{jg}  y_{m+1-i,g}(-1)^{j+1}T^{(s)}_{g,j}\alpha_j = \sum_{g}  y_{m+1-i,g}(-1)^{g}A^{(g)}(h).
\end{align}
Let us justify the last equality. Recall that
\begin{align}
    T^{(s)}_{g,j} =\frac{1-(-1)^{g+j}}{(g+j)2^{g+j}}, 
\end{align}
which will be different from $0$ if and only if $g+j = 1$ mod $2$, or in other words, $j+1 = g$ mod $2$. Thus we can substitute the $(-1)^{j+1}$ with a $(-1)^{g}$, obtaining that $y_{i,g} = (-1)^g y_{m+1-i,g}$. Second, recall that $A^{(g)}(h)$ are either odd or even functions of $h$, if $g$ is even or odd, respectively. As such, the inclusion of higher-order terms not included in the truncation of $A^{(g)}(h)$ to order $2s$ will respect the time anti-symmetry. One example of this behavior can be found in Eqs. 41 and 42 in~\cite{blanes2006fourth}, for the fourth order commutator-free quasi-Magnus operators. 

Finally, the quadrature operator will also ultimately respect the time anti-symmetry in~\eqref{eq:time_symmetry_parameter_constraint}. The odd rows in the quadrature matrix $Q^{(s,k)}$,~\eqref{eq:Gauss_Legendre_quadrature}, multiplied by $A^{(k)}$ with even $k$, will exhibit symmetric behavior for nodes symmetrically chosen around $t_{\frac{1}{2}}$. Conversely, the even rows, corresponding to odd $k$, will symmetrically change sign across $t_{\frac{1}{2}}$ due to the sign change in $c_k$ and the symmetry of weights $w_k$ across $t_{\frac{1}{2}}$. Together with the symmetry in Trotter product formulas, this indicates that the commutator-free Magnus product formulas presented here fulfill the requirements to use the multi-product formula machinery presented in~\cite{low2019well}. Alternative constructions of commutator-free quasi-Magnus operators also respect the time anti-symmetry indicated here, see for instance ~\cite[Eq. 34]{alvermann2011high}.  
It is worth mentioning that the Suzuki formalism is also amenable to multi-product formulas~\cite{wiebe2010higher,watkins2022time,low2019well}.

\subsection{Computing the error of multi-product formulas}
So far we have discussed why we can use multi-product formulas to decrease the error of CFQMs. Here we explain how to bound their error, thus being able to compute the step size $h$ that will result in an equal or lower bound than the target error.
We follow a methodology similar to the one used in Theorem 2 in~\cite{low2019well}. 

As a first step, we need to bound the error contributions of order greater or equal to $O(h^{2s'+1})$ in the product formula. Propositions~\ref{prop:expMagnus_Taylor}-\ref{prop:Quadrature} already display the adequate format for this: we only have to start the summatory from order $O(h^{2s'+1})$ instead of $O(h^{2s+1})$. In contrast, the error bound we used for the Trotter-Suzuki product formula step is not useful here. Instead, we use the result from Lemma F.2 in the supplementary material in~\cite{childs2018toward}, which bounds the Taylor remainder of $\exp(\lambda)$ of order $k$, $\mathcal{R}_k(\exp(\lambda))$, by
\begin{equation}
    \|\mathcal{R}_k(\exp(\lambda))\|\leq \frac{|\lambda|^{k+1}}{(k+1)!}\exp(|\lambda|).
\end{equation}

We can use this Lemma to compute the $O(h^{2s'+1})$ error of each exponential in $S_{2s}\left(\overline{U}_m^{[2s]}\left( h\right)\right)$, so in our case $\lambda = \sum_{k=1}^{s}z_{i,k} A_k h$. Then we leverage~\cref{eq:error_exponentials}, obtain the Trotter error of $O(h^{2s'+1})$. Adding up all the errors we obtain the $O(h^{2s'+1})$ error for the CFQM.

The second step is converting such an error bound into an error bound for the multi-product formula. First, we use again~\cref{eq:error_exponentials} to bound the error of multiplying together $k_j$ product formulas of time length $h/k_j$. Then, we take into account the final linear combination of unitaries, multiplying each term by its corresponding $|a_j|$, and adding them up. As stated above, the result is an upper bound to the multi-product formula error as a function of $h$, $a_j$ and $k_j$. Then, we can maximize $h$ subject to the error target we imposed.

\bibliographystyle{plainnat}
\bibliography{main}

\end{document}